\documentclass[a4paper,11pt]{article}

\usepackage{amsmath}
\usepackage{amsfonts}
\usepackage{amssymb}         
\usepackage{amsopn}
\usepackage{amsthm}
\usepackage{graphicx}
\usepackage{epsfig}
\usepackage{times}
\usepackage{verbatim}
\usepackage{mathtools}
\usepackage{subcaption}
\usepackage{authblk}
\usepackage{hyperref}
\usepackage[left=1.0in,top=1.2in,right=1.0in]{geometry}
\usepackage{enumerate}
\usepackage{algorithm}
\usepackage{algorithmic}
\usepackage{rotating}
\usepackage{hyperref}
\usepackage{qcircuit}
\usepackage{color}
\usepackage{lineno}
\usepackage{afterpage}
\usepackage{longtable}

\title{Leveraging Secondary Storage to Simulate Deep 54-qubit Sycamore
  Circuits}

\author[1]{Edwin Pednault\thanks{Corresponding author; \url{pednault@us.ibm.com}}}
\author[1]{John A.~Gunnels}
\author[1]{Giacomo Nannicini}
\author[1]{Lior Horesh}
\author[1]{Robert Wisnieff}
\affil[1]{IBM T.J.~Watson Research Center, Yorktown Heights, NY}

\date{}

\begin{document}


\maketitle


\begin{abstract}
  In a recent paper, we showed that secondary storage can extend the
  range of quantum circuits that can be practically simulated with classical
  algorithms. Here we refine those techniques and apply them to the
  simulation of Sycamore circuits with $53$ and $54$ qubits,
  with the entanglement pattern ABCDCDAB that has proven difficult to
  classically simulate with other approaches. Our analysis shows that
  on the Summit supercomputer at Oak Ridge National Laboratories, such
  circuits can be simulated with high fidelity to arbitrary depth in a
  matter of days, outputting all the amplitudes.
\end{abstract}

\section{Introduction}
There has been tremendous progress in the construction of quantum
computers with superconducting qubits
\cite{castelvecchi2017quantum}. As the hardware progresses, it is
increasingly difficult to classically simulate the circuits that can
be executed on existing chips, a crucial task to -- among other things
-- verify that the degree to which hardware is behaving as
expected. The literature contains several papers that discuss this
task and that have been published or posted online in the past two
years
\cite{pednault2017breaking,boixo2017simulation,chen201864,li2018quantum,chen2018classical,markov2018quantum,villalonga2018flexible,chen2019teleport,villalonga2019frontier,guo2019PEPS,zhang2019classical}. Here
we extend the analysis on the use of secondary storage initially
reported in our previous work \cite{pednault2017breaking}. As we
argued in that paper, secondary storage can extend the computational
reach of supercomputers for the simulation of quantum circuits --- an
idea initially suggested by \cite{haner2017simulation}.  We estimate
that on the Summit supercomputer at Oak Ridge National Laboratories,
secondary storage allows the simulation of $53$- and $54$-qubit
Sycamore circuits \cite{rieffel2019sycamore} with high fidelity to
arbitrary depth. The Sycamore circuits are a direct descendant of the
``universal random circuits'' described in
\cite{boixo2018supremacy}. In particular, for 20 cycles of the
entanglement pattern ABCDCDAB, which is specifically designed to
challenge classical simulation algorithms, we estimate that the
computations would take approximately two and a half days. While we
did not carry out these computations, we provide a detailed
description of the proposed simulation strategy as well as the time
estimation methodology, which is based on published results and on
internal benchmarks. The main building blocks of our approach are the
same as those that we discussed in \cite{pednault2017breaking},
namely: exploitation of separable gates via a hyperedge representation
of the tensor network; allowing contractions between non-adjacent
tensors; tensor slicing; and sporadic read/write operations to
access/store slices of the quantum state in secondary storage.

The rest of this paper is organized as follows. In
Sect.~\ref{sec:tensor} we provide a review of the main building blocks
of our simulation strategy. Sect.~\ref{sec:sycamore} describes the
class of circuits studied in this work. Sect.~\ref{sec:strategy} gives
a detailed explanation of the simulation strategy and its use of
secondary storage. Sect.~\ref{sec:times} concludes the paper by
estimating the time required by the proposed simulations, and
discussing the methodology to compute such estimates. The Appendix
contains a detailed listing of the operations performed by the
proposed simulation strategy.

\section{Brief overview of tensor contraction deferral}
\label{sec:tensor}
The simulation algorithm that we propose is based on the idea of
partitioning a quantum circuit into subcircuits that can be simulated
independently, at the expense of extra bookkeeping to account for
entanglement between subcircuits. We ensure that the final results are
correct by appropriately recombining the different subcircuits and, in
some sense, ``resolving'' the entanglement. Rather than insisting that
all subcircuits reside in primary storage, i.e., RAM, we allow for storing
the results of some of the calculations on secondary storage, e.g.,
disk. This is particularly effective when combined with slicing
techniques, which further partition the quantum state by iteratively
fixing the value of some of the indices in the tensor network.

We now give a brief overview of the main components of our simulation
strategies; for details and several examples, we refer the reader to
our earlier work \cite{pednault2017breaking}. A {\em tensor} is a
multilinear map with a set of indices to address its elements. In the
context of this paper, each index takes value in $\{0,1\}$. As
discussed in \cite{pednault2017breaking}, a {\em tensor network} is a
hypergraph $G = (V, E)$ such that each node is associated with a
tensor and each hyperedge with an index of the adjacent
tensors. Hyperedges between nodes represent shared indices that must
be summed over.  A summation over shared indices is called a {\em
  contraction}. A tensor $A_{i_1,\dots,i_m,j_1,\dots,j_m}$ is {\em
  diagonal} if it is nonzero only if $i_k = j_k$ for $k=1,\dots,m$. A
tensor is {\em separable} if it can be obtained from a diagonal tensor
with a permutation, i.e., there exist functions $f_1,\dots,f_m$ such
that $A_{f_1(j_1,\dots,j_m),\dots,f_m(j_1,\dots,j_m),j_1,\dots,j_m}$
is diagonal. Our hypergraph representation is designed to take
advantage of separable tensors, since the computational resources
necessary to perform a contraction between several tensors can be
significantly reduced in the presence of indices shared among multiple
tensors --- represented by hyperedges. Given a quantum circuit, we can
construct a tensor network by letting the gates correspond to tensors,
and the qubit lines roughly correspond to the indices (i.e. edges and
hyperedges). To simulate large circuits, we rely extensively on {\em
  contraction deferral} and {\em tensor slicing}, which we describe more fully below.

Contraction deferral is a technique first introduced in
\cite{aaronson2016complexity}. Its use in large-scale simulations was
pioneered in \cite{pednault2017breaking}. Contraction deferral is
defined as the contraction of arbitrary sets of (potentially
non-adjacent) tensors in the tensor network; this is in contrast with
the adjacent contraction discussed in the seminal work
\cite{markov2008simulating} and in several subsequent papers, e.g.,
\cite{boixo2017simulation}.  A deferred contraction performs the usual
summation over shared indices (i.e., edges interior to the set being
contracted), and applies an outer product to the non-shared
indices. As it is a generalization of the traditional adjacent
contraction, contraction deferral opens up new simulation strategies
that can lead to reduced memory requirements. In particular, we
partition the tensor network into sub-hypergraphs corresponding to
subcircuits, each of which includes fewer qubits than the initial
circuit. Within each circuit we perform computations following the
so-called ``Schr\"odinger approach'' \cite{aaronson2016complexity},
i.e., evolving the full quantum state of the subcircuit by applying
layers of gates one at a time. To do so, we must use contraction
deferral whenever we apply tensors corresponding to entangling gates
between different subcircuits.

Tensor slicing is the idea of iterating over several instances of a
circuit in which certain hyperedges (i.e., tensor indices) are fixed
to one of their possible values. While this does not necessarily
reduce the number of operations to be performed, it allows reordering
the computations so that they take place in ways that are potentially
more efficient. This is particularly crucial when using secondary
storage, which is slower than primary storage and must therefore be
used sparingly; with tensor slicing, we reorganize the calculations so
that only a few selected slices (rather than full tensors) reside in
primary storage at any given time. Our scheme extends the simulation
strategy in \cite{haner2017simulation}: we choose a set of indices,
slice them by looping over every possible combination of their values,
and use a superset of those qubits to efficiently organize and address
information located in secondary storage.


\begin{figure}[tbp]
  \centering
  \includegraphics[width=1.0\textwidth]{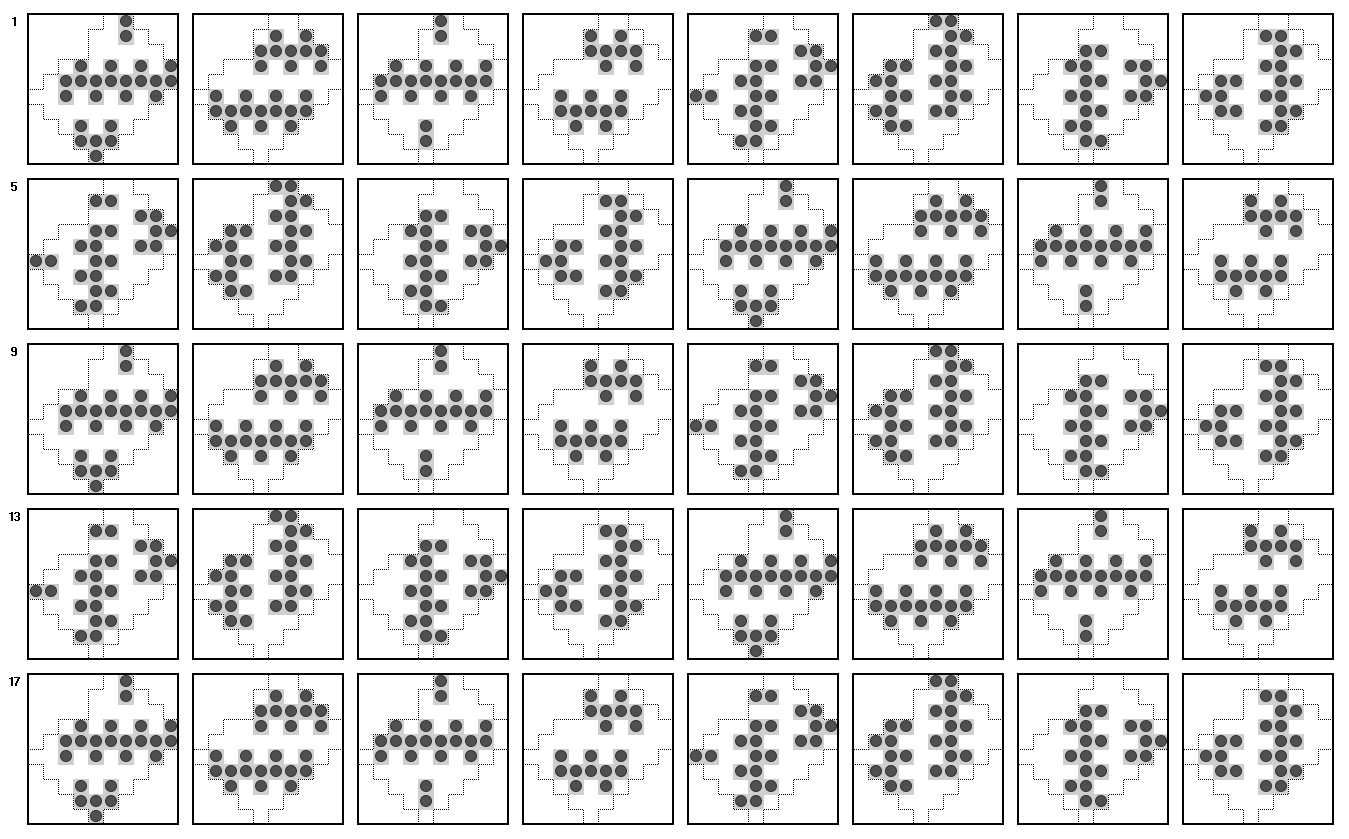}
  \caption{Gate pattern for a 20-cycle, 53-qubit, Sycamore ABCDCDAB
    circuit.  Single-qubit gates are merged into their neighboring
    two-qubit gates, and the two-qubit gates in each cycle are
    partitioned into two layers for illustration purposes to make the
    individual gates easy to identify. These
    transformations result in the 40-layer circuit depicted.
    Dots and shading are used to identify which pairs of qubits are
    being operated upon.
  }
  \label{fig:53_gatepattern}
\end{figure}

\begin{figure}[tbp]
  \centering
  \includegraphics[width=1.0\textwidth]{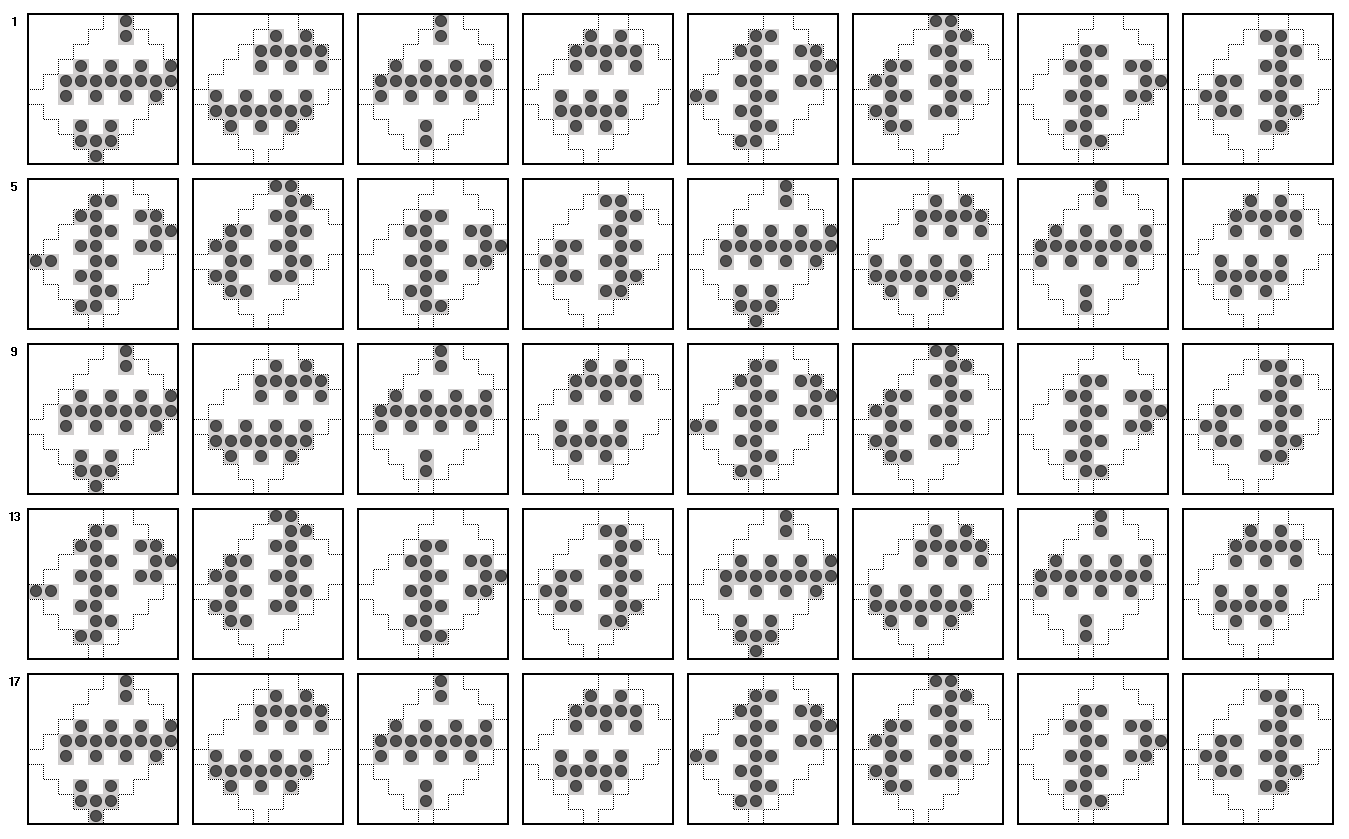}
  \caption{Gate pattern for a 20-cycle, 54-qubit, Sycamore ABCDCDAB
    circuit.  Single-qubit gates are merged into their neighboring
    two-qubit gates, and the two-qubit gates in each cycle are
    partitioned into two layers for illustration purposes to make the
    individual gates easy to identify.  These transformations result
    in the 40-layer circuit depicted.
    Dots and shading are used to identify which pairs of qubits are
    being operated upon.
}
  \label{fig:54_gatepattern}
\end{figure}

\section{Sycamore circuits}
\label{sec:sycamore}
A recent paper \cite{rieffel2019sycamore} describes a new class of
random quantum circuits consisting of alternating layers of
single-qubit gates and two-qubit gates.  The combination of a layer of
single-qubits gates followed by a layer of two-qubit gates is referred
to in~\cite{rieffel2019sycamore} as a {\em cycle}. Our paper discusses
the classical simulation of circuits of this class; hence, we describe
them here in more detail. Gates are applied to all qubits in each
single-qubit gate layer, and to almost all qubits in each two-qubit
gate layer. The two-qubit gates are all non-diagonal, non-separable
and their unitary representation varies as a function of both the
location of the gate within the qubit layout, and the depth at which
the gate is applied in the circuit.  The former reflects variations in
gate tuning, while the latter reflects variations in pulse
synchronization over time.  Circuits consist of several cycles of
single-qubit gates followed by two-qubit gates, together with a final
layer of single-qubit gates.  The single-qubit gates are randomly
selected from the set $\{\sqrt{X}, \sqrt{Y}, \sqrt{W}\}$.  The
two-qubit gates implement the following unitary:
\begin{equation}
  \begin{bmatrix}
    1 & 0 & 0 & 0 \\
    0 & e^{i(\Delta_+ + \Delta_- )}\cos\theta
      & -ie^{i(\Delta_+ - \Delta_{-,\textrm{off}} )}\sin\theta
      & 0 \\
    0 & -ie^{i(\Delta_+ + \Delta_{-,\textrm{off}} )}\sin\theta
      & e^{i(\Delta_+ - \Delta_- )}\cos\theta
      & 0 \\
    0 & 0 & 0 & e^{i(2\Delta_+ - \phi)} \\
  \end{bmatrix}
  ,
\end{equation}
where $\theta$ and $\phi$ are nominally $90^{\circ}$ and $30^{\circ}$,
respectively, and where $\Delta_+$, $\Delta_-$, and
$\Delta_{-,\textrm{off}}$ are detuning terms. For simulation purposes,
all single-qubit gates can be aggregated with neighboring two-qubit
gates, yielding an equivalent circuit consisting of (potentially
unique) two-qubit gates only. The method for selecting single-qubit
gates ensures that these two-qubit unitary operations are also
randomized. The difficulty of simulation is therefore determined entirely
by the pattern of two-qubit gates in the circuit.

Figs.~\ref{fig:53_gatepattern} and~\ref{fig:54_gatepattern} illustrate
the ``ABCDCDAB'' patterns of two-qubit gates used in the 53- and
54-qubit random circuits described in~\cite{rieffel2019sycamore} to
test the Sycamore quantum device.  This pattern is intentionally
devised to make the resulting circuits difficult to simulate
classically.  For illustration purposes, each cycle of two-qubit gates
is depicted as two layers in the figures, so that we can unambiguously
indicate the pairs of qubits involved in each gate operation (i.e.,
vertical pairs versus horizontal pairs).  Thus, the first two layers
of two-qubit gates illustrated correspond to the ``A'' cycle, the next
two layers to the ``B'' cycle, and so on. In this representation, the
first row corresponds to an ``ABCD'' sequence of cycles, the second
row corresponds to a ``CDAB'' sequence of cycles, and the five rows
illustrated in Figs.~\ref{fig:53_gatepattern}
and~\ref{fig:54_gatepattern} correspond to the 20-cycle circuits
generated according to the ABCDCDAB rules described
in~\cite{rieffel2019sycamore}.

\begin{figure}[tbp]
  \centering
  \includegraphics[width=1.0\textwidth]{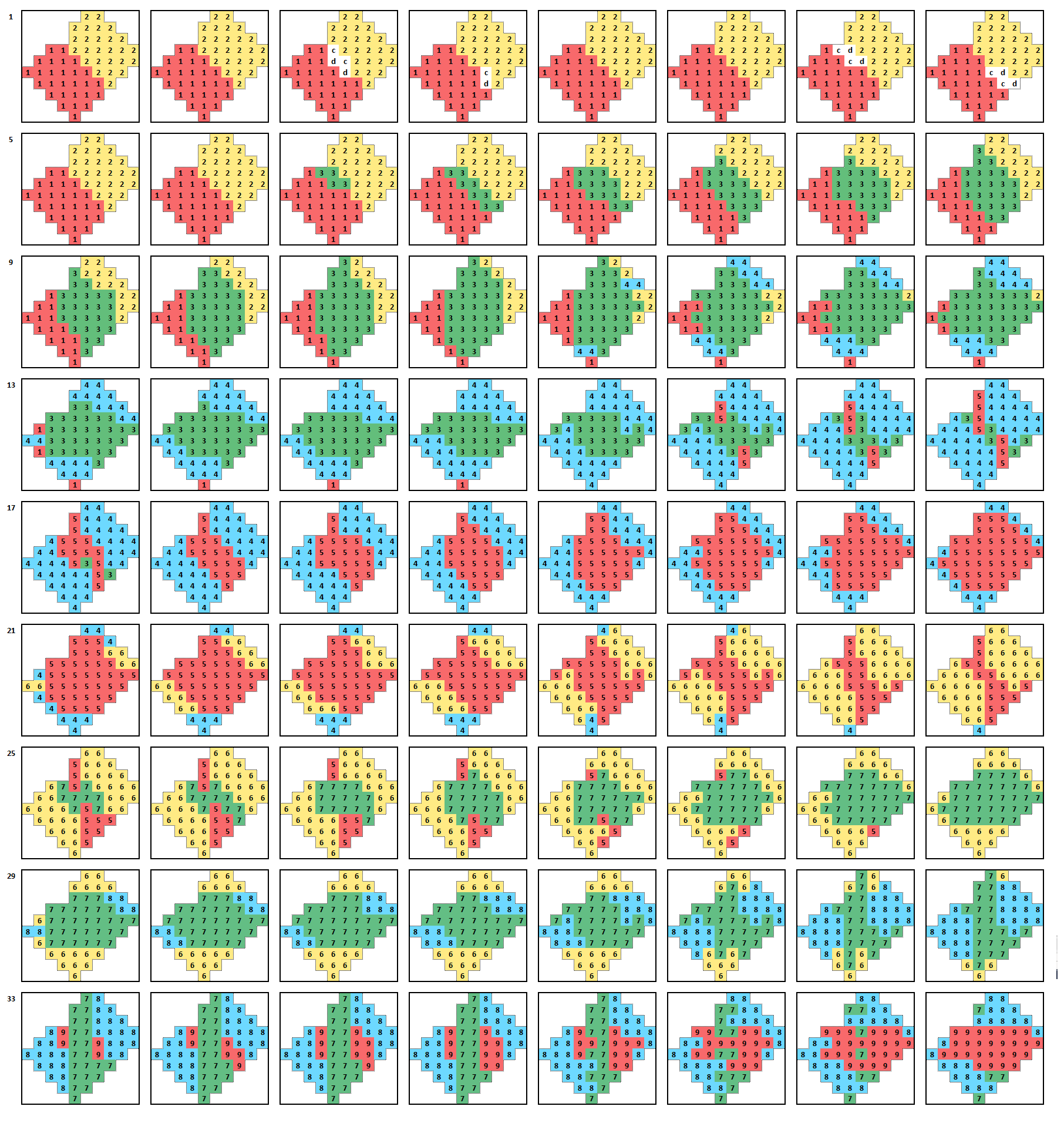}
  \caption{Partitioning of a 36-cycle, 53-qubit, Sycamore ABCDCDAB
    circuit to leverage secondary storage.  Numbers and colors are used to
    indicate regions of gates within the circuit that are grouped together
    to form subcircuits, and also to refer to specific subcircuits in the text.
    Contraction deferral is applied to the gates labeled ``cd.''
  }
  \label{fig:53_partitioning}
\end{figure}

\begin{figure}[tbp]
  \centering
  \includegraphics[width=1.0\textwidth]{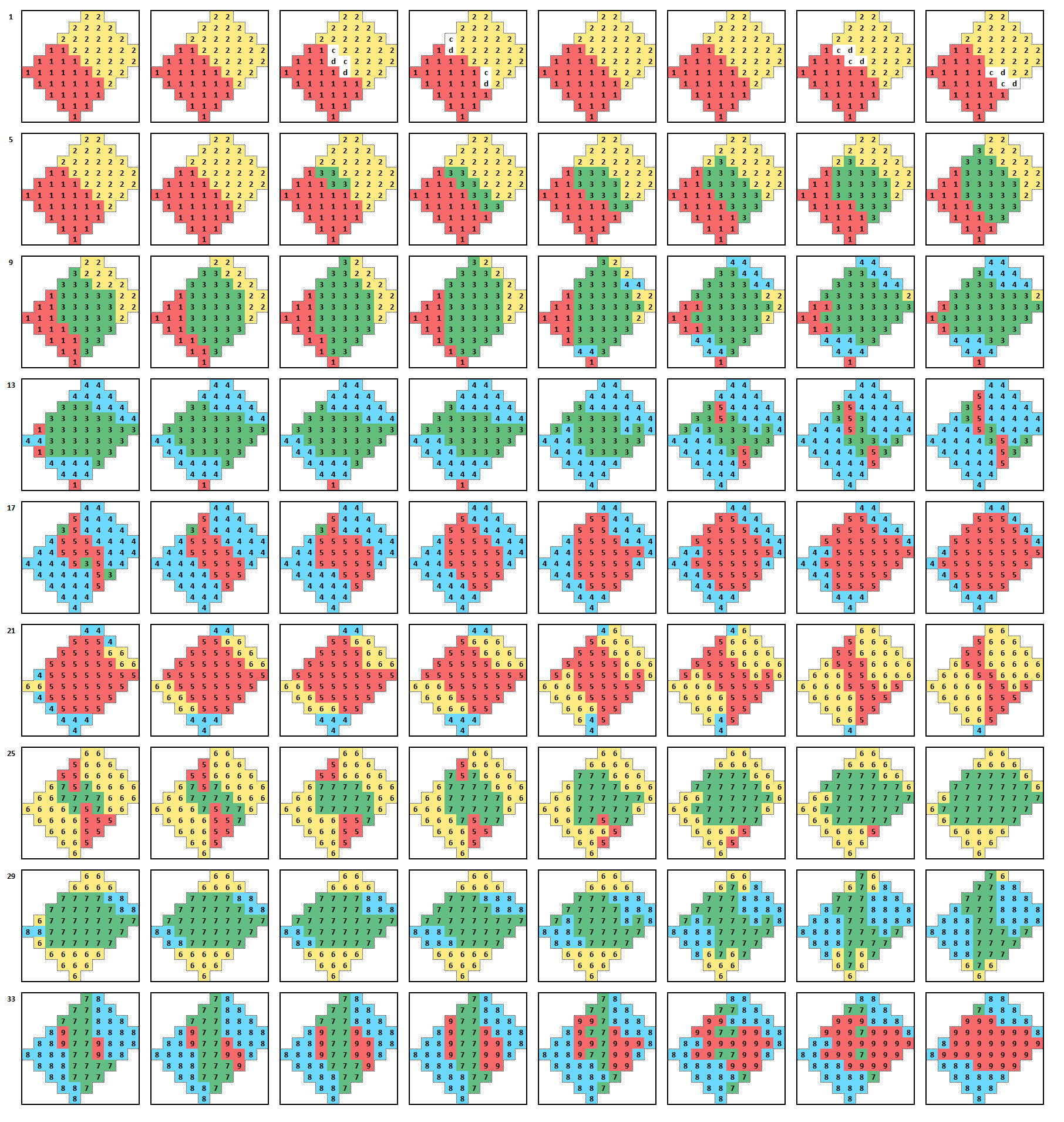}
  \caption{Partitioning of a 36-cycle, 54-qubit, Sycamore ABCDCDAB
    circuit to leverage secondary storage.  Numbers and colors are used to
    indicate regions of gates within the circuit that are grouped together
    to form subcircuits, and also to refer to specific subcircuits in the text.
    Contraction deferral is applied to the gates labeled ``cd.''
  }
  \label{fig:54_partitioning}
\end{figure}

\section{Proposed simulation strategy}
\label{sec:strategy}
In \cite{haner2017simulation} the authors suggest that solid-state
disk, or more generally secondary storage, could be used to supplement
main memory in order to simulate circuits whose quantum states are too
large to store in main memory alone.  In~\cite{pednault2017breaking},
we combined our in-memory methods with those of
\cite{haner2017simulation}, describing a viable computation scheme
that exploits secondary storage to simulate deeper circuits than was
thought possible. We now apply the approach discussed in
\cite{pednault2017breaking} to 53- and 54-qubit Sycamore circuits,
showing in this section a computation scheme that allows their
simulation on an existing supercomputer, Summit. The cost of such a
scheme is discussed in Sect.~\ref{sec:times}.

The simulation method in \cite{haner2017simulation} can be seen as a
tensor slicing approach. Qubits (and the corresponding tensor indices)
are divided into ``global'' qubits, which are sliced and used to
address across processing nodes, and ``local'' qubits, corresponding
to tensor indices used to address tensor slices stored on each
processing node.  In \cite{haner2017simulation}, circuits are
partitioned so that all gates within a subcircuit can be applied to
the local slice of the quantum state, without communicating quantum
state information among processing nodes. Such zero-communication
updates of a local quantum state are possible when all non-diagonal
gates in a subcircuit are applied to local qubits only. They are
also possible for a handful of additional circumstances described in
\cite{haner2017simulation}.  In effect, circuits are partitioned by
selecting different subsets of local qubits and analyzing which gates
can be applied to them without communication. This determines the
subcircuits. During simulation, communication between processing nodes
occurs only when the simulation switches from one subcircuit to
another. When a communication phase takes place, the memory layout of
quantum state tensors is reorganized so that different global and
local qubits (i.e., tensor indices) are selected, according to the
subcircuits that have to be simulated in the subsequent phase.

\begin{figure}[tb]
  \centering
  \includegraphics[width=0.9\textwidth]{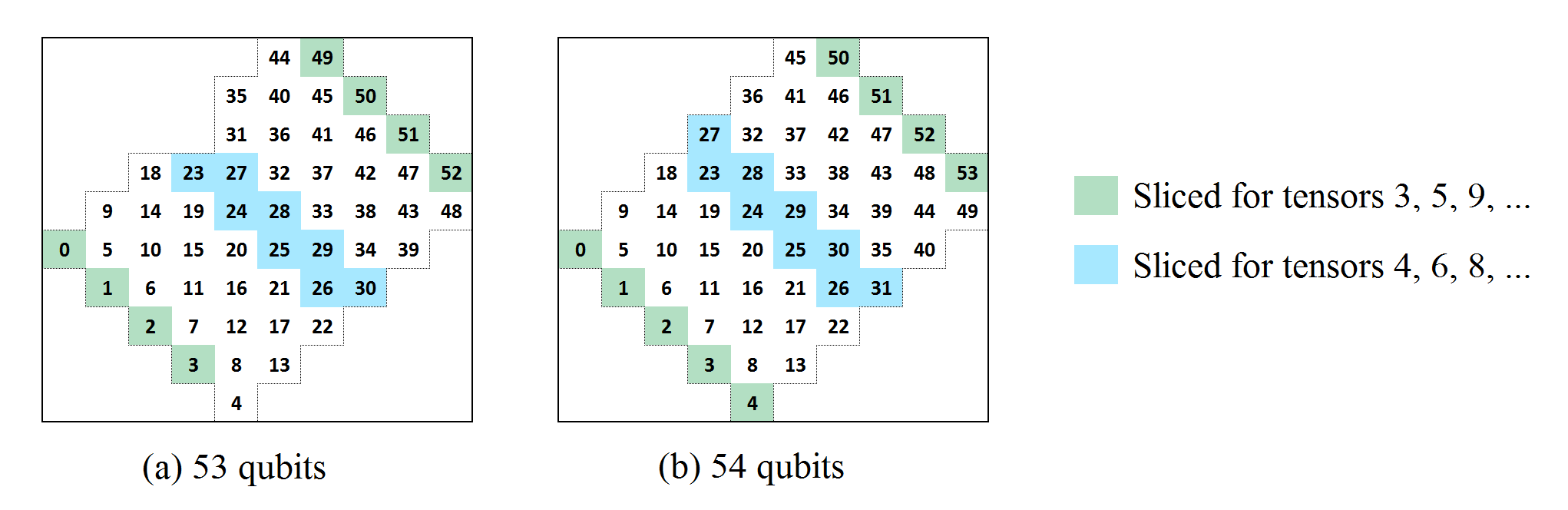}
  \caption{Qubit numbering scheme and first-level tensor slicing
    strategy for 53- and 54-qubit Sycamore circuits.}
  \label{fig:53_54_slicing}
\end{figure}

The in-memory method that we presented in \cite{pednault2017breaking}
considers circuit partitionings in which the resulting tensors either
fit in available aggregate primary memory in their entirety, or their
slices can be computed using available primary memory (using other
tensors already computed and stored in primary memory). With this
approach, the resulting tensors and/or their slices will generally be
larger than the primary memories of individual processing nodes; this
represents a difference between \cite{pednault2017breaking} and
\cite{haner2017simulation}.

As discussed in \cite{pednault2017breaking}, we combine the
zero-communication strategies of \cite{haner2017simulation} with our own
tensor partitioning strategy to leverage
secondary storage when quantum states are too large to fit in
aggregate primary memory.  Because secondary storage is typically
orders of magnitude slower than main memory, the viability of using it
depends on the extent to which the number of read/write cycles can be
minimized or overlapped with computation.  To this end, we first
employ the in-memory methods of~\cite{pednault2017breaking}, aiming to
maximize the number of gates that can be simulated using available
aggregate memory; the resulting quantum state is calculated in slices
and written to secondary storage.  The partitioning methods discussed
in \cite{haner2017simulation} can then be applied to the remaining
gates in the circuit, setting the number of ``local'' qubits according
to the size of aggregate memory, rather than the memory sizes
available on individual processing nodes. This increases the size of
the resulting tensor slices, allowing the application of many more
gates to the local quantum state before additional secondary storage
read/write cycles are needed. The resulting subcircuits can be further
partitioned into sub-subcircuits, using the methods
of~\cite{haner2017simulation}, to minimize internode communication in
the overall calculations. We now provide details about these
partitionings for the specific circuits studied in this paper.

Figs.~\ref{fig:53_partitioning} and~\ref{fig:54_partitioning}
illustrate the first level of circuit partitioning for 36-cycle
Sycamore ABCDCDAB circuits with 53 and 54 qubits, respectively. In
this first phase, we use the in-memory methods
of~\cite{pednault2017breaking} to simulate the subcircuits~1 and~2
illustrated in these figures, performing tensor contraction deferral
on the gates labeled ``cd.''  The outer-most qubits of the resulting
tensors are used as ``global'' qubits, with the corresponding slices
contracted in order to allow the simulation of subcircuit~3, slice
by slice. Each resulting slice for subcircuit~3 is written to disk.
Fig.~\ref{fig:53_54_slicing} illustrates the ``global'' qubits that
are sliced in this first phase of simulation.  In the case of the
53-qubit circuit, qubits~0--3 and~49--52 are sliced in the simulation
of subcircuit~3; for the 54-qubit circuit, qubits~0--4 and~50--53 are
sliced.

In the second phase, qubits~23--30 are sliced for the 53-qubit circuit
and qubits~23--31 are sliced for the 54-qubit circuit.  In both cases,
the following steps are performed for each slice: the slice is read
from disk, the gates in subcircuit~4
shown in Figs.~\ref{fig:53_partitioning} and~\ref{fig:54_partitioning}
for the respective circuits are applied, and the slice is written back
to disk.

This process is repeated for each subsequent subcircuit, with the
choice of sliced qubits alternating between those used for subcircuit
3 and those used for subcircuit~4.  Specifically, for subcircuits 5, 7
and 9, in the 53-qubit circuit we slice qubits~0--3 and~49--52, while
in the 54-qubit circuit we slice qubits~0--4 and~50--53.  For
subcircuits 6 and 8, in the 53-qubit circuit we slice qubits 23--30,
while in the 54-qubit circuit we slice qubits 23--31.  As with
subcircuit~4, slices are read from disk, processed, and then written
back to disk.

\begin{figure}[tb]
  \centering
  \includegraphics[width=0.9\textwidth]{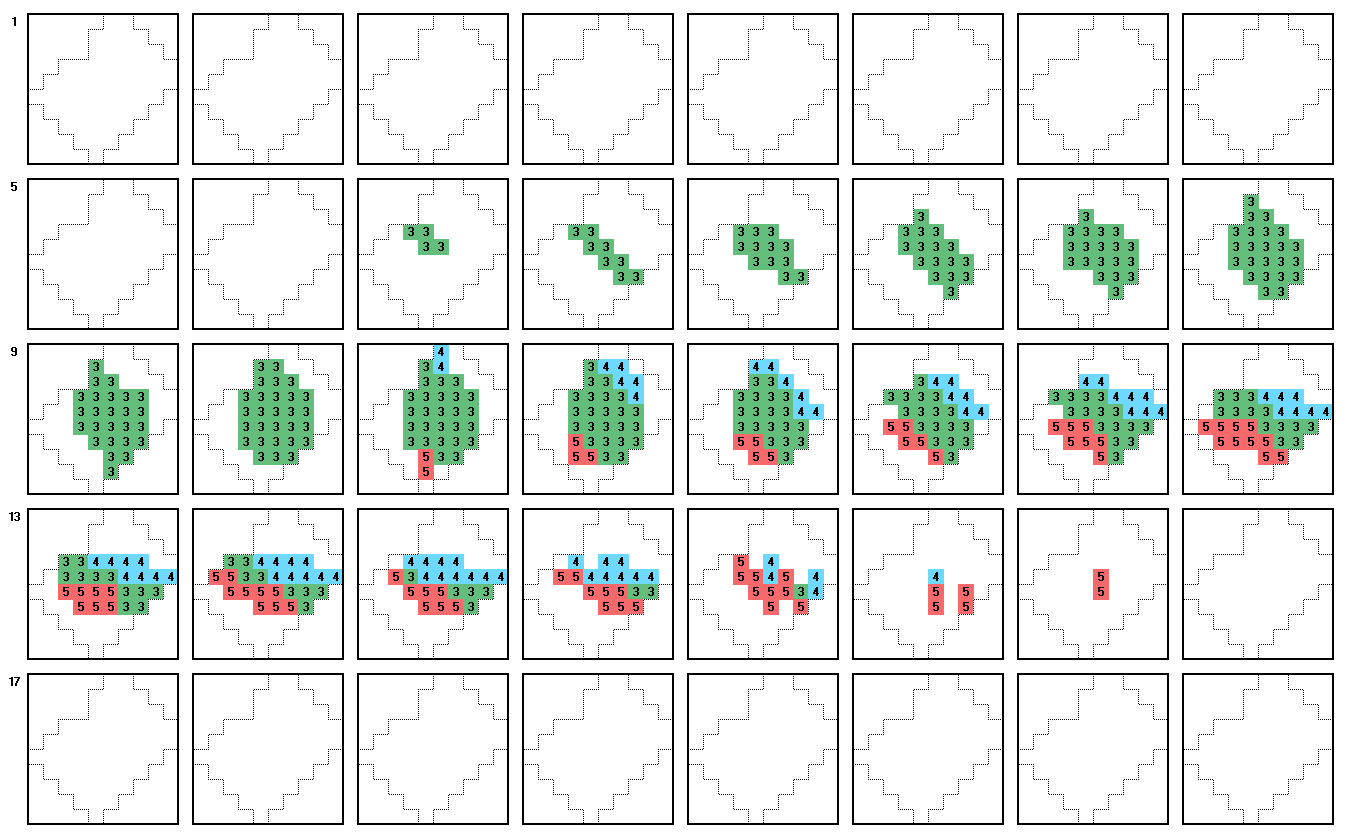}
  \caption{Partitioning of subcircuit 3 to minimize
    all-to-all communication for the 20-cycle, 53-qubit Sycamore
    circuit shown in Fig.~\ref{fig:53_gatepattern}.}
  \label{fig:53_tensor_3}
\end{figure}

\begin{figure}[tb]
  \centering
  \includegraphics[width=0.9\textwidth]{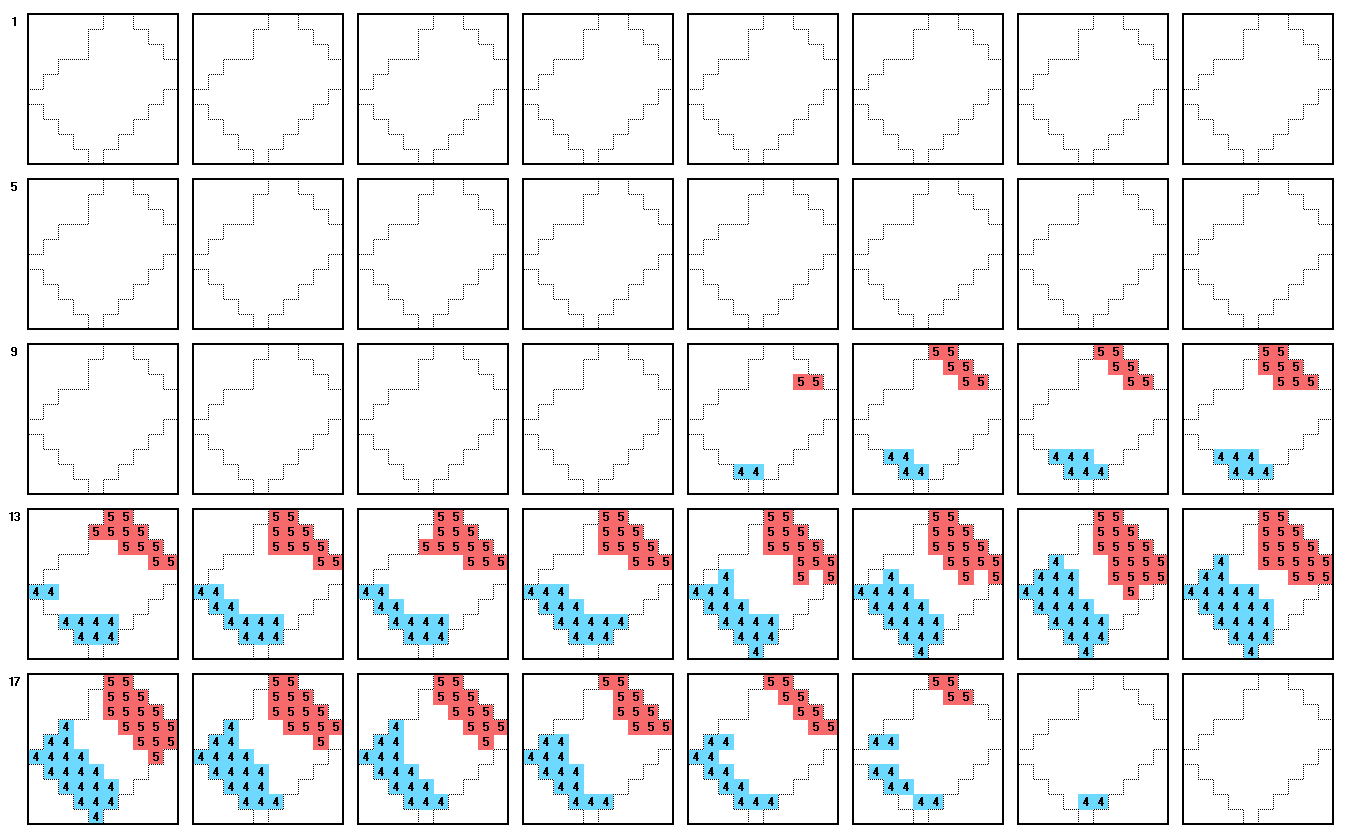}
  \caption{Partitioning of subcircuit 4 to minimize
    all-to-all communication for the 20-cycle, 53-qubit Sycamore
    circuit shown in Fig.~\ref{fig:53_gatepattern}.}
  \label{fig:53_tensor_4}
\end{figure}

\begin{figure}[tb]
  \centering
  \includegraphics[width=0.9\textwidth]{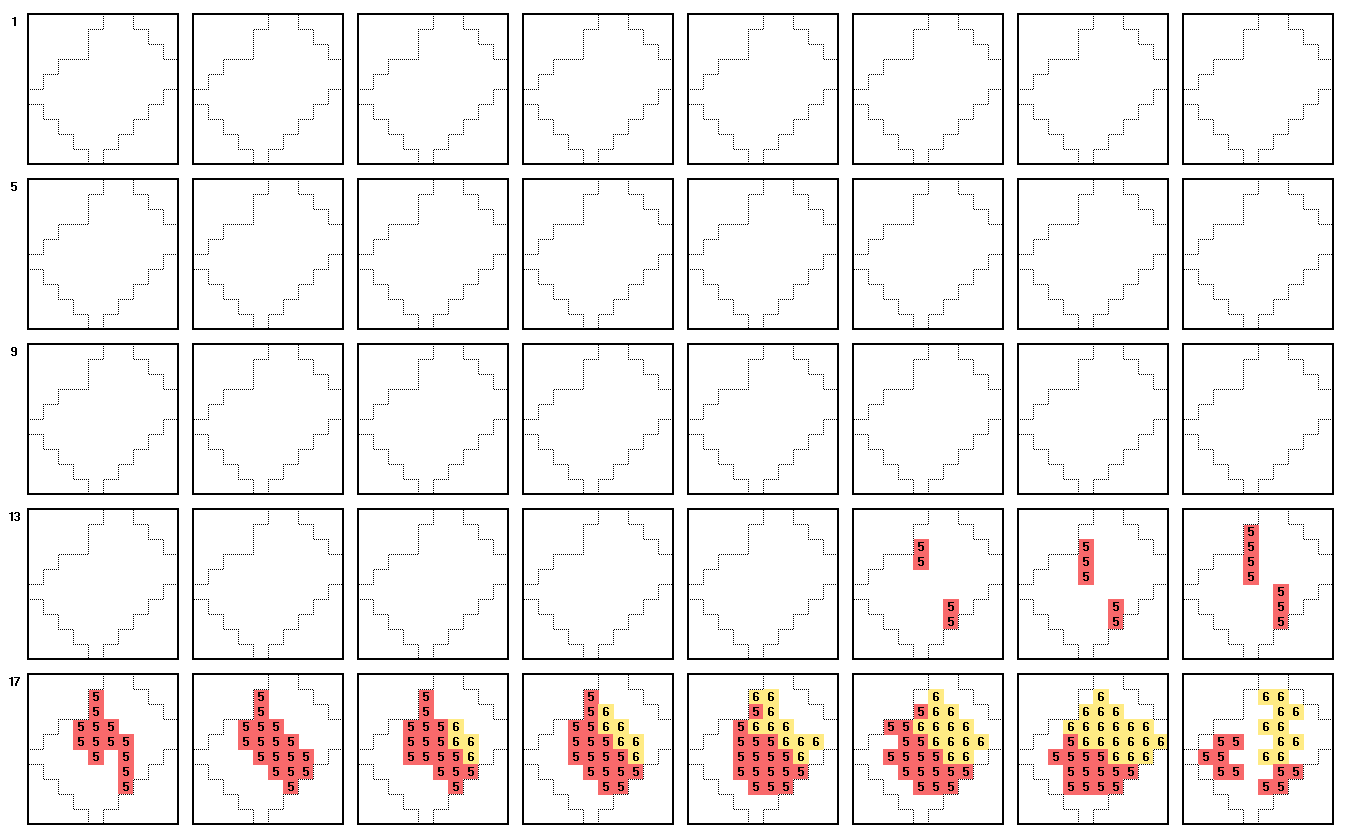}
  \caption{Partitioning of subcircuit 5 to minimize
    all-to-all communication for the 20-cycle, 53-qubit Sycamore
    circuit shown in Fig.~\ref{fig:53_gatepattern}.}
  \label{fig:53_tensor_5}
\end{figure}

\begin{figure}[tb]
  \centering
  \includegraphics[width=0.9\textwidth]{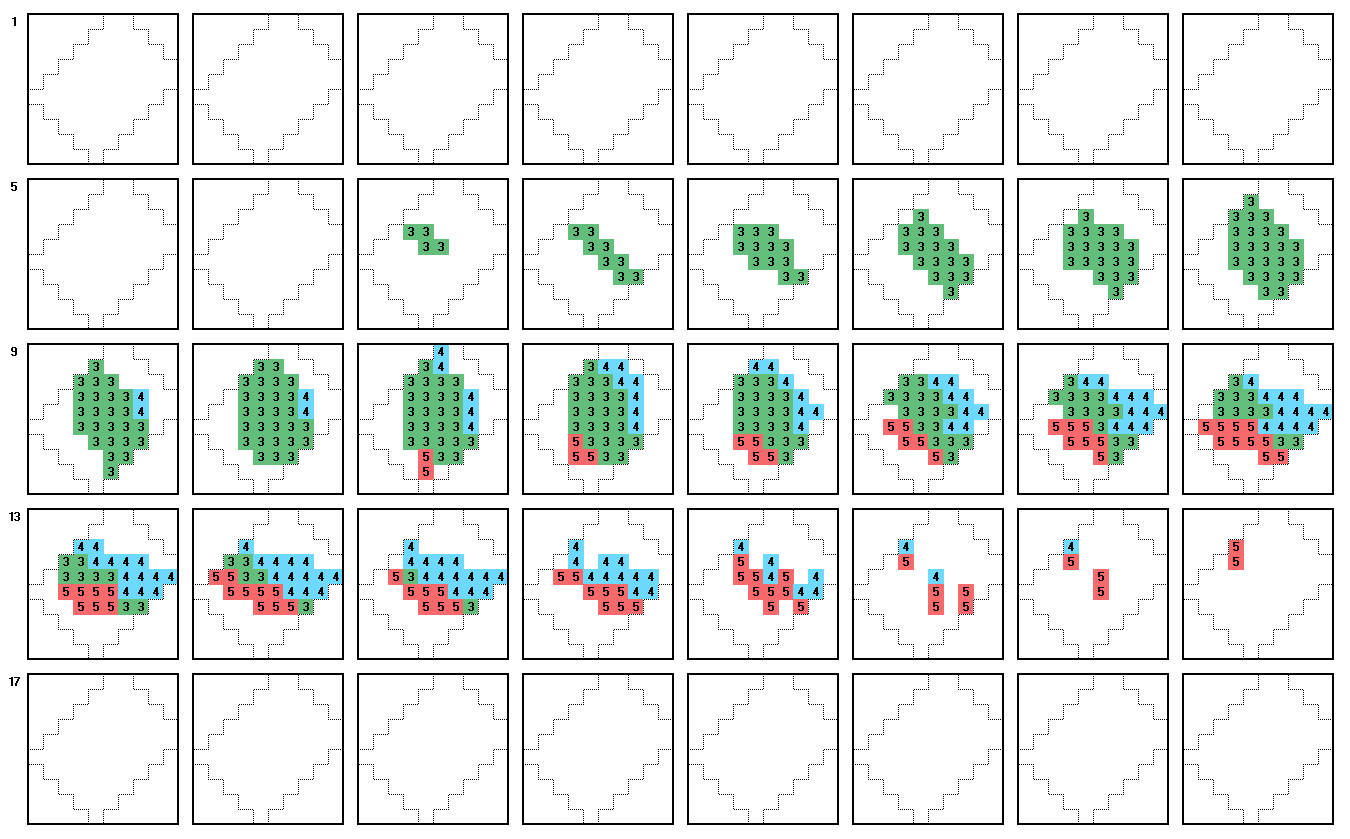}
  \caption{Partitioning of subcircuit 3 to minimize
    all-to-all communication for the 20-cycle, 54-qubit Sycamore
    circuit shown in Fig.~\ref{fig:54_gatepattern}.}
  \label{fig:54_tensor_3}
\end{figure}

\begin{figure}[tb]
  \centering
  \includegraphics[width=0.9\textwidth]{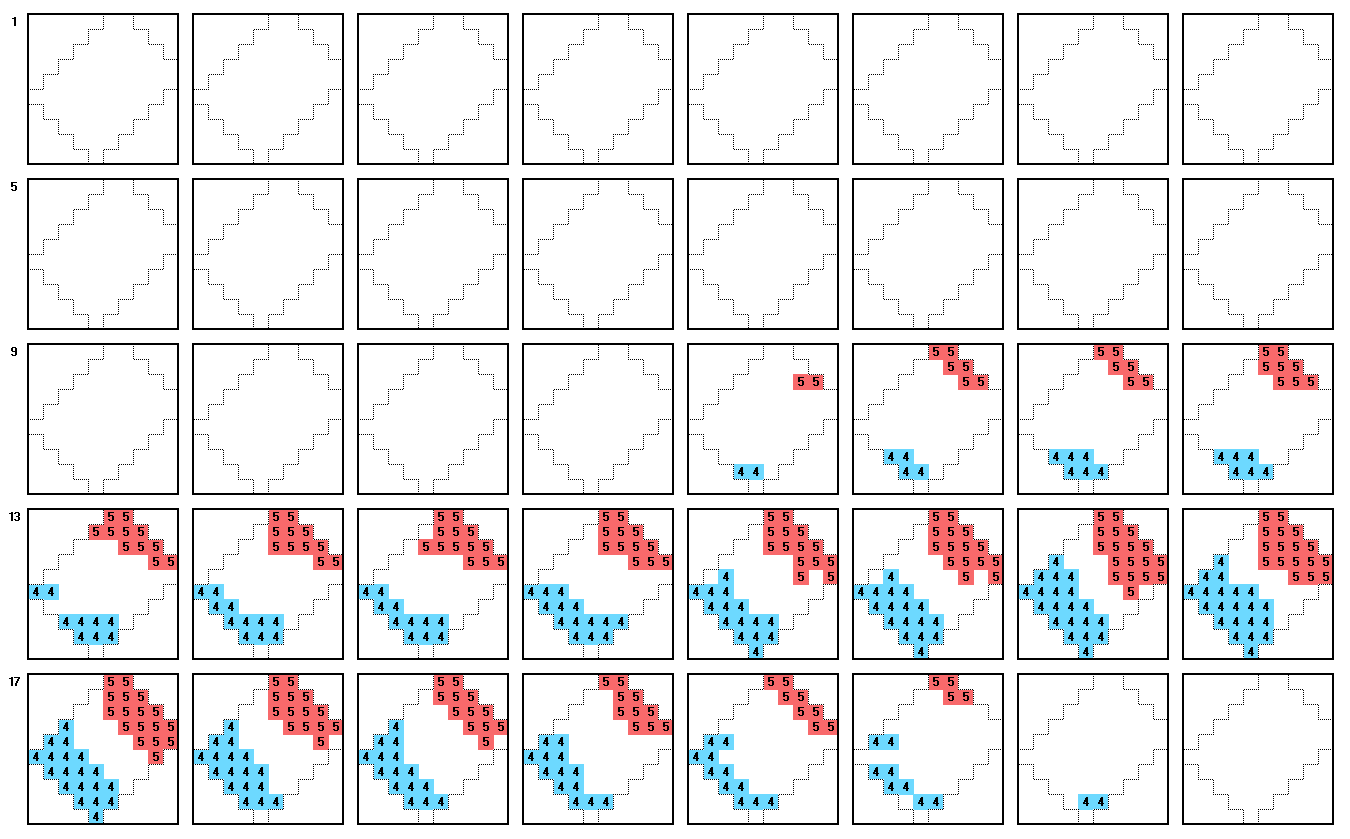}
  \caption{Partitioning of subcircuit 4 to minimize
    all-to-all communication for the 20-cycle, 54-qubit Sycamore
    circuit shown in Fig.~\ref{fig:54_gatepattern}.}
  \label{fig:54_tensor_4}
\end{figure}

\begin{figure}[tb]
  \centering
  \includegraphics[width=0.9\textwidth]{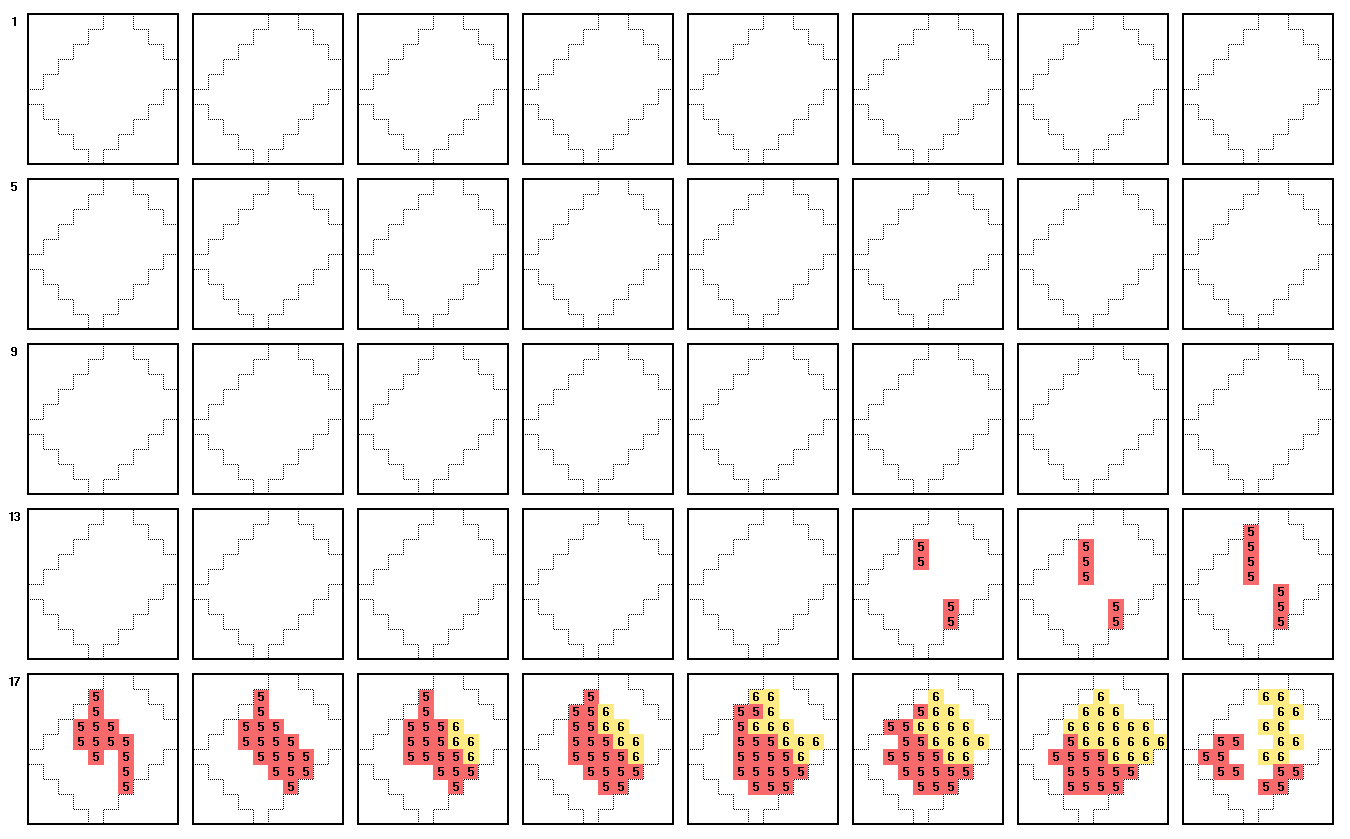}
  \caption{Partitioning of subcircuit 5 to minimize
    all-to-all communication for the 20-cycle, 54-qubit Sycamore
    circuit shown in Fig.~\ref{fig:54_gatepattern}.}
  \label{fig:54_tensor_5}
\end{figure}

To ensure that we efficiently transfer data to/from secondary storage,
we organize the data on secondary storage as $2^{16}$ logical files
for the 53-qubit circuit, and $2^{18}$ logical files for the 54-qubit
circuit.  In the case of the 53-qubit circuit, files are indexed by the
values of qubits 0--3, 23--30, and 49--52; each logical file contains
$2^{37}$ complex amplitudes corresponding to qubits 4--22 and 31--48.
In the case of the 54-qubit circuit, files are indexed by the values of
qubits 0--4, 23--31, and 50--53; each logical file contains $2^{36}$
complex amplitudes corresponding to qubits 5--22 and 32--49.  Thus, in
the first phase of simulation for the 53-qubit circuit (i.e., the phase
in which tensor 3 is written to disk), we write 256 logical files to
secondary storage for each of the 256 values of qubits 0--3 and 49--52
that are being sliced; these files correspond to the 256 possible
values of qubits 23--30.  For 54-qubit circuits, 512 logical
files are written to secondary storage for each of the 512 values of
qubits 0--4 and 50--53 that are being sliced; these files correspond
to the 512 possible values of qubits 23--31.

In the second phase of simulation of the 53-qubit circuit (i.e., the
phase in which subcircuit 4 is simulated), for each of the 256 values
of qubits 23--30 that are being sliced, we read 256 logical files from
secondary storage, corresponding to the 256 possible values of qubits
0--3 and 49--52.  Once these 256 files of amplitudes are loaded into
memory, we apply the gates in subcircuit 4 and write each updated
slice back to storage. Similarly, for the 54-qubit circuit, for each
of the 512 values of qubits 23--31 that are being sliced, we read 512
logical files from secondary storage, corresponding to the 512
possible values of qubits 0--4 and 50--53.  Updated slices are written
back to secondary storage as 512 files of amplitudes.  These access
patterns are repeated for each subsequent phase of processing.  The
above approach guarantees that individual logical files are always read
or written in their entirety, and they are never read or written multiple
times in a single read or write cycle. Access overhead per
read/write cycle is thereby minimized.

The above slicing strategy is designed to minimize the number of disk
accesses by maximizing the number of ``local'' qubits employed in each
disk slice, which is 45 qubits for both the 53- and the 54-qubit
circuit.  As discussed earlier, the slicing methodology
in~\cite{haner2017simulation} is applied recursively to these 45-qubit
slices, to minimize the number of all-to-all communication steps that
must be performed in order to simulate subcircuits 3--9, shown in
Figs.~\ref{fig:53_partitioning} and~\ref{fig:54_partitioning}.
Figs.~\ref{fig:53_tensor_3}--\ref{fig:54_tensor_5} illustrate these
recursive partitionings for tensors 3, 4, and 5 in the case of
20-cycle circuits.  These partitionings correspond to slicing an
additional 13 qubits (i.e., in addition to the qubits sliced for disk
access purposes), in order to distribute work across 4096~nodes and
across each pair of IBM Power~9 sockets within those nodes. We employ
a socket-level slicing strategy to enable each socket to work
independently, and to avoid Non-Uniform Memory Access (NUMA) overhead
when memory accesses cross socket boundaries.

As shown in Figs.~\ref{fig:53_tensor_3} and~\ref{fig:54_tensor_3},
subcircuit 3 is recursively partitioned into three sub-subcircuits,
labeled 3, 4, and 5.  For the 53-qubit circuit, the sub-subcircuit
labeled 3 is sliced on qubits 4--10 and 43--48, and for the 54-qubit
circuits we slice qubits 5--10 and 43--49.  These specific qubits are
selected so that the corresponding slices of tensors 1 and 2, which
are small, can be pre-distributed across sockets; this way the
contractions needed to start simulating subcircuit 3 can be performed
in-place without communication.  For 53-qubit circuits, the
sub-subcircuit labeled 4 is sliced on qubits 4--16, and, for 54-qubit
circuits, on qubits 5--17.  This redistribution requires an all-to-all
exchange of amplitudes across sockets.  For 53-qubit circuits, the
sub-subcircuit labeled 5 is sliced on qubits 36--48, and, for 54-qubit
circuits, on qubits 37--49, again requiring all-to-all communication.

As shown in Figs.~\ref{fig:53_tensor_4} and~\ref{fig:54_tensor_4},
subcircuit 4 is recursively partitioned into two sub-subcircuits
labeled 4 and 5.  For 53-qubit circuits, the sub-subcircuit labeled 4
is sliced on qubits 40--52, and for 54-qubit circuits on qubits
41--53.  The sub-subcircuit labeled 5 is sliced on qubits 0--12 for
both 53- and 54-qubit circuits.

As shown in
Figs.~\ref{fig:53_tensor_5} and~\ref{fig:54_tensor_5}, subcircuit 5 is
recursively partitioned into two sub-subcircuits labeled 5 and 6.  For
53-qubit circuits, the sub-subcircuit labeled 5 is sliced on qubits
36--48, and, for 54-qubit circuits, on qubits 37--49.  The
sub-subcircuit labeled 6 is sliced on qubits 4--16 for 53-qubit
circuits, and on qubits 5--17 for 54-qubit circuits.

\section{Estimated running times}
\label{sec:times}
We estimate running times for the above simulation strategy on the
Summit supercomputer using a combination of published performance
figures and early IBM internal benchmarks.

Because we directly employ the partitioning strategy of
\cite{haner2017simulation} in a recursive fashion, and the
resulting 45-qubit disk slices coincide with the 45-qubit circuits
simulated in~\cite{haner2017simulation}, we use the performance
figures in~\cite{haner2017simulation} to estimate per-disk-slice
computational costs. This implicitly assumes that we are directly
using the implementation described in~\cite{haner2017simulation} for
the computations. Since \cite{haner2017simulation} employs 8,192 nodes
of the Cori II supercomputer, we can more easily extrapolate predicted
performance across a corresponding 8,192 sockets on Summit.

\begin{table}[tbp]
  \scriptsize
  \centering
  \begin{tabular}{|*{10}{c|}}
    \hline
    & Disk  & All-to- & 5Q & Tensor &      & Contrac- & &  & \\
    & trasfers & alls & kernels & ranks &     & tion & Compute & \% of &   \\
    & per disk & per disk & per disk & per & Num & cost tot. & time & total & Achieved \\
    Tensor & slice & slice & slice & socket & gates & FLOPs & (days) & time & PFLOPS \\
    \hline
    1 &    & 0.000977 & 28 & 28 & 84 &    & 0.002082 & 0.08\% & 0.0308 \\
    2 &    & 0.000977 & 25 & 27 & 84 &    & 0.001859 & 0.07\% & 0.0173 \\
    Contraction &  &  &    & 31 &    & 1.181$\cdot 10^{21}$ & 0.117058 & 4.59\% & 116.7304 \\
    3.3 &    &        & 16 & 32 & 63 &    & 0.010658 & 0.42\% & 18.4865 \\
    3.4 &    &  1     &  6 & 32 & 23 &    & 0.003997 & 0.16\% & 17.9975 \\
    3.5 &    &  1     &  8 & 32 & 26 &    & 0.005329 & 0.21\% & 15.2587 \\
    Disk write  & 1 & 1 &  & & & & & & \\
    Disk read  & 1 & 1 &  & & & & & & \\
    4.4 &    &        & 11 & 32 & 49 &    & 0.007327 & 0.29\% & 20.9141 \\
    4.5 &    &  1     & 10 & 32 & 45 &    & 0.006661 & 0.26\% & 21.1275 \\
    Disk write  & 1 & 1 &  & & & & & & \\
    Disk read  & 1 & 1 &  & & & & & & \\
    5.5 &    &        &  9 & 32 & 35 &    & 0.005995 & 0.24\% & 18.2583 \\
    5.6 &    &  1     &  7 & 32 & 21 &    & 0.004663 & 0.18\% & 14.0850 \\
    Disk write  & 1 & 1 &  & & & & & & \\
    \hline
    Subtotals &   &   &  & & & & & & \\
    Compute &   &   & 120 &   &   & 1.181$\cdot 10^{21}$ & 0.165631 & 6.50\% & 87.4462 \\
    All-to-alls &   & 9.001953 &   &   &   &   & 0.487725 & 19.13\% &   \\
    Disk I/O &  5 &    &   &   &   &   & 1.896296 & 74.37\% &   \\
    \hline
    Total &  5  & 9.001953 & 120 & 32.67243 & 430 &   & 2.549652 & 100.00\% & 87.4462 \\
    \hline
  \end{tabular}

  \caption{Running time estimates to simulate the 20-cycle, 53-qubit
    Sycamore circuit shown in Fig.~\ref{fig:53_gatepattern}.  Tensors
    3.3, 3.4, and 3.5 correspond to the partitionings of subcircuit
    3 shown in Fig.~\ref{fig:53_tensor_3}, tensors 4.4 and 4.5 to the
    partitionings of subcircuit 4 shown in Fig.~\ref{fig:53_tensor_4},
    and tensors 5.5 and 5.6 to the partitionings of subcircuit 5 shown
    in Fig.~\ref{fig:53_tensor_5}.  The number of 5-qubit kernels is
    the number of aggregated gates spanning no more than 5 qubits,
    created by grouping gates together within each subcircuit.  The
    contraction cost is the total number floating-point operations
    needed to perform the tensor contractions associated with
    entanglement indices, when tensors 1 and 2 are contracted in
    preparation for simulating subcircuit 3.  The tensor ranks per
    socket indicate the sizes of the corresponding tensors in terms of
    the effective number of local qubits per socket.  For entries
    above the subtotal line, the compute times are either the
    estimated times to perform gate operations based on the number of
    5-qubit kernels, or the estimated time to perform the tensor 1 and
    2 contractions, depending on the row in the table.  Entries below
    the subtotal line factor in the costs of performing
    all-to-all communication and disk I/O. In the case of gate
    operations, the achieved PetaFLOPs per second column is the total
    number of floating-point operations
    without gate aggregation divided by the estimated compute time
    with gate aggregation.}
  \label{fig:53_20_runtime}

\end{table}

\begin{table}[tbp]
  \scriptsize
  \centering
  \begin{tabular}{|*{10}{c|}}
    \hline
    & Disk  & All-to- & 5Q & Tensor &      & Contrac- & &  & \\
    & trasfers & alls & kernels & ranks &     & tion & Compute & \% of &   \\
    & per disk & per disk & per disk & per & Num & cost tot. & time & total & Achieved \\
    Tensor & slice & slice & slice & socket & gates & FLOPs & (days) & time & PFLOPS \\
    \hline
    1 &    & 0.001953 & 28 & 30 & 84 &    & 0.004164 & 0.07\% & 0.0616 \\
    2 &    & 0.001953 & 26 & 30 & 87 &    & 0.003867 & 0.07\% & 0.0687 \\
    Contraction &  &  &    & 33 &    & 9.445$\cdot 10^{21}$ & 0.936466 & 16.14\% & 116.7304 \\
    3.3 &    &        & 15 & 32 & 59 &    & 0.019984 & 0.42\% & 18.4865 \\
    3.4 &    &  1     &  8 & 32 & 31 &    & 0.010658 & 0.18\% & 18.1931 \\
    3.5 &    &  1     &  8 & 32 & 27 &    & 0.010658 & 0.18\% & 15.8456 \\
    Disk write  & 1 & 1 &  & & & & & & \\
    Disk read  & 1 & 1 &  & & & & & & \\
    4.4 &    &        & 11 & 32 & 49 &    & 0.014655 & 0.25\% & 20.9141 \\
    4.5 &    &  1     & 10 & 32 & 45 &    & 0.013323 & 0.23\% & 21.1275 \\
    Disk write  & 1 & 1 &  & & & & & & \\
    Disk read  & 1 & 1 &  & & & & & & \\
    5.5 &    &        &  9 & 32 & 37 &    & 0.011990 & 0.21\% & 19.3016 \\
    5.6 &    &  1     &  7 & 32 & 21 &    & 0.009326 & 0.16\% & 14.0850 \\
    Disk write  & 1 & 1 &  & & & & & & \\
    \hline
    Subtotals &   &   &  & & & & & & \\
    Compute &   &   & 122 &   &   & 9.445$\cdot 10^{21}$ & 1.035091 & 17.84\% & 107.2342 \\
    All-to-alls &   & 9.003906 &   &   &   &   & 0.975661 & 16.81\% &   \\
    Disk I/O &  5 &    &   &   &   &   & 3.792593 & 65.35\% &   \\
    \hline
    Total &  5  & 9.003906 & 122 & 33.80735 & 440 &   & 5.803345 & 100.00\% & 107.2342 \\
    \hline
  \end{tabular}

  \caption{Running time estimates to simulate the 20-cycle, 54-qubit
    Sycamore circuit shown in Fig.~\ref{fig:54_gatepattern}.  Tensors
    3.3, 3.4, and 3.5 correspond to the partitionings of subcircuit
    3 shown in Fig.~\ref{fig:54_tensor_3}, tensors 4.4 and 4.5 to the
    partitionings of subcircuit 4 shown in
    Fig.~\ref{fig:54_tensor_4}, and tensors 5.5 and 5.6 to the
    partitionings of subcircuit 5 shown in
    Fig.~\ref{fig:54_tensor_5}. The number of 5-qubit kernels is
    the number of aggregated gates spanning no more than 5 qubits,
    created by grouping gates together within each subcircuit. The
    contraction cost is the total number floating-point operations
    needed to perform the tensor contractions associated with
    entanglement indices, when tensors 1 and 2 are contracted in
    preparation for simulating subcircuit 3.  The tensor ranks per
    socket indicate the sizes of the corresponding tensors in terms of
    the effective number of local qubits per socket.  For entries
    above the subtotal line, the compute times are either the
    estimated times to perform gate operations based on the number of
    5-qubit kernels, or the estimated time to perform the tensor 1 and
    2 contractions, depending on the row in the table.  Entries below
    the subtotal linefactor in the costs of performing
    all-to-all communication and disk I/O.  In the case of gate
    operations, the achieved PetaFLOPs per second column is the total
    number of floating-point operations
    without gate aggregation divided by the estimated compute time
    with gate aggregation.}
    \label{fig:54_20_runtime}
\end{table}

To account for the differences between the gate set
of~\cite{haner2017simulation} and Sycamore circuits, we use the
following two facts: in~\cite{haner2017simulation}, gates are
aggregated together into $k$-qubit kernels represented by $2^k \times
2^k$ unitary matrices; and amplitudes are updated using matrix-matrix
and/or matrix-vector
calculations. The gate aggregation effectively normalizes computations
across gate sets, making them independent from the details of
individual gates. For simulations performed on Cori II, gate
aggregation in \cite{haner2017simulation} uses $k_{max} = 5$, with
actual kernel sizes sometimes being less than 5~qubits depending on
the aggregated gates. To leverage the performance figures reported
in~\cite{haner2017simulation}, we therefore perform the same form of
gate aggregation on Sycamore circuits. 
Tables~\ref{fig:53_20_runtime} and~\ref{fig:54_20_runtime} summarize
the results obtained with these gate aggregations for the 53- and
54-qubit circuits illustrated in Figs.~\ref{fig:53_gatepattern}
and~\ref{fig:54_gatepattern}, respectively.  In these tables, the
``5-qubit kernels per disk slice'' column identifies the number of
aggregate gates constructed for the corresponding subcircuit
or sub-subcircuit.

After gate aggregation, we estimate execution times for gate
operations using Tables 1 and 2 in~\cite{haner2017simulation}.
Specifically, we use the ``Time'' and ``Comm.'' columns of Table 2
in~\cite{haner2017simulation} to estimate computation time from total
execution time, by factoring out the reported percentage of
communication and synchronization time. We then divide the computation
times by the number of aggregate gates (i.e., clusters) listed in
Table 1 in~\cite{haner2017simulation} to obtain overall average
execution times per aggregate gate.  These averages can be used to
estimate execution times for arbitrary numbers of aggregate gates,
assuming simulations are performed on Cori II.  To obtain
corresponding time estimates for Summit, we scale the estimates by the
ratio of the High Performance Linpack (HPL) benchmark figure for Cori
II (14,014.70 TeraFLOPs/sec) versus Summit (148,600.00 TeraFLOPs/sec);
this accounts for the substantially greater performance of Summit when
performing, e.g., the matrix-vector calculations entailed by the use
of gate aggregation.  The calculations for 45-qubit simulations yield
an expected execution time of 2.38380 seconds per aggregate gate on
Cori II, and 0.22482 seconds per aggregated gate on Summit.  The
runtime estimates shown in Tabs.~\ref{fig:53_20_runtime}
and~\ref{fig:54_20_runtime} for tensors 3, 4 and 5 are obtained by
multiplying the number of aggregate gates in each of their
sub-subcircuits by 0.22482 seconds, and further multiplying by the
number of disk slices per tensor (i.e., 256 slices for the 53-qubit
circuit and 512 slices for the 54-qubit circuit).  Tensors 1 and 2, on
the other hand, represent 30-qubit calculations or less; therefore, we
use the 30-qubit performance figures from Tables 1 and 2
in~\cite{haner2017simulation} to obtain an estimated 0.025097 seconds
per aggregate gate on Summit for these tensors.

The ``achieved FLOPs per second'' columns in
Tabs.~\ref{fig:53_20_runtime} and~\ref{fig:54_20_runtime} provide a
sanity check for the above time estimates.  For rows corresponding to
gate operations, this column reports the number of floating-point
operations that would be performed without gate aggregation divided by
the estimated execution times.  As such, the resulting figures provide
an indication of the implied efficiency of the time estimates. As can
be seen, the time estimates for gate operations yield results that are all near or below
11\% of the 191~PetaFLOPs/sec peak double-precision performance
expected across 8,192 sockets. Therefore, there is room to potentially
improve upon these estimates by leveraging the capabilities of
Summit's NVIDA GPUs: this would allow the simulation of individual
gate operations, without resorting to gate aggregation, and the use of
cuBLAS routines to implement the corresponding matrix-vector
operations.

To obtain time estimates for the contractions of tensors 1 and 2, we
use the performance figures reported in Table 1
in~\cite{villalonga2019frontier}.  The simulation method presented
in~\cite{villalonga2018flexible,villalonga2019frontier} employs the
``bristle-brush'' strategy outlined in~\cite{pednault2017blog}.  A key
characteristic of this simulation strategy is that computations are
dominated by very large tensor contractions across many tensor indices
simultaneously.  Consequently, we directly use the performance figures
reported in Table 1 in~\cite{villalonga2019frontier} to estimate
contraction times.  Because we assume double-precision calculations,
as in~\cite{haner2017simulation}, we convert the performance figures
in~\cite{villalonga2019frontier} from single-precision to
double-precision. To do so, we multiply the single-precision
computation rates of~\cite{villalonga2019frontier} by the ratio
between the double-precision (7.8 TeraFLOPs/sec) and single-precision
(15.7 TeraFLOPs/sec) peak performance rate of Summit's NVIDIA GPUs.
Performing this calculation and taking the worst case yields as
estimated 14.249 TeraFLOPs per socket (116.73 PetaFLOPs/sec for 8,192
sockets). We use this rate in Tabs.~\ref{fig:53_20_runtime}
and~\ref{fig:54_20_runtime} to estimate execution times for the
contraction of tensors 1 and 2.

We estimate all-to-all and disk I/O times using results from early
IBM internal benchmarks.  These
benchmarks indicate that a network injection rate of 7 GB/sec per node
(3.5 GB/sec per socket) should be easily achieved during an all-to-all
across the entire machine.  This figure represents $\approx 30\%$ of
the 23 GB/sec per node peak injection rate (11.5 GB/sec per socket)
that characterizes the bisection bandwidth of Summit.  The maximum
reported file-system transfer rate is 2.2 TB/sec for random-access I/O
(2.5 TB/sec for pure sequential I/O). We assume that all disk storage
operations use single precision, while in-memory calculations use
double precision. Thus, the estimates in Tabs.~\ref{fig:53_20_runtime}
and~\ref{fig:54_20_runtime} are based on a transfer rate of 2~TB/sec
and a storage density of 8 bytes per amplitude (i.e., single-precision
complex). Benchmark tests suggest that allocating only a subset of
nodes to the task of performing disk I/O can be more efficient because
it may avoid contention; those nodes then become the distribution
points to the rest of the system when spreading computations across a
majority of the nodes.  Tabs.~\ref{fig:53_20_runtime}
and~\ref{fig:54_20_runtime} model this arrangement by incorporating an
all-to-all communication cost for every disk read or write operation.

The resulting estimated running times are summarized in
Tabs.~\ref{fig:53_20_runtime} and~\ref{fig:54_20_runtime}. As these
tables show, with the performance model discussed in this section we
obtain an overall estimate of 2.55 days to compute all $2^{53}$
amplitudes of a 20-cycle, 53-qubit, Sycamore ABCDCDAB circuit with all
amplitudes stored on disk, and 5.80 days for the corresponding
54-qubit circuit.  To store amplitudes on disk in single precision,
64~PiB of disk space are required for 53-qubit circuits, and 128~PiB for
54-qubit circuits.  Both fit within the 250~PiB available on Summit.

\begin{figure}[tb]
  \centering
  \includegraphics[width=0.9\textwidth]{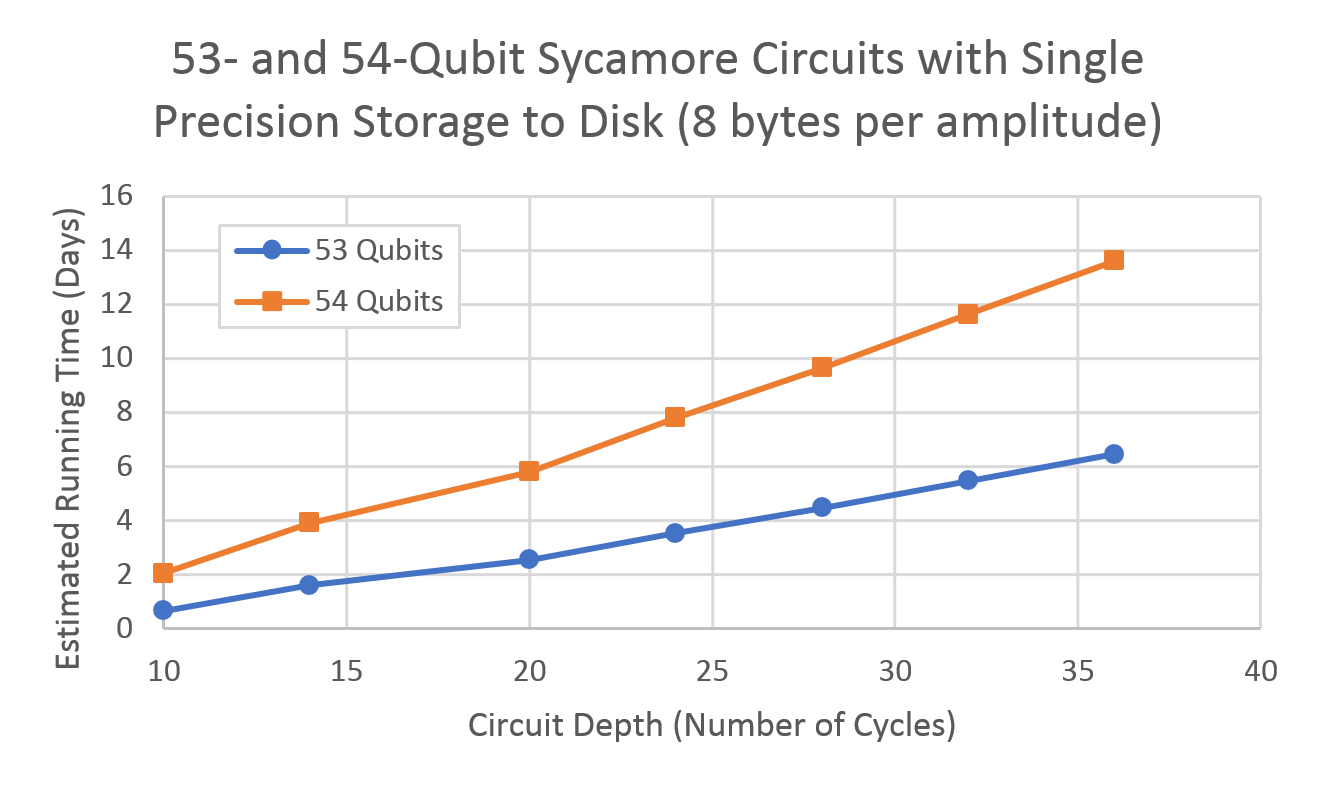}
  \caption{Graph of total runtime estimates for fully simulating both
    53- and 54-qubit, Sycamore ABCDCDAB circuits of various depths,
    with all amplitudes calculated and stored on disk.}
  \label{fig:53_54_combined}
\end{figure}

The above analysis can be repeated for all depths suggested by
Figs.~\ref{fig:53_partitioning} and~\ref{fig:54_partitioning}; i.e.,
10, 14, 20, 24, 28, 32, and 36 cycles.  Doing so yields the results
reported in Tables~\ref{tab:53_manycycles}
and~\ref{tab:54_manycycles}, and plotted in
Fig.~\ref{fig:53_54_combined}.  As these tables and figure illustrate,
estimated execution times grow linearly with the depth of the
circuits. We remark that the required disk space remains constant,
because with the above approach there is a maximum number of slices
that are stored on disk at any given time. Thus, the disk occupation
is 64~PiB for 53-qubit circuits and 128~PiB for 54-qubit circuits,
regardless of the number of cycles.

\begin{table}[tbp]
  \centering
  \begin{tabular}{|c|c|c|c|c|}
    \hline
    Number & Disk Xfers & All-to-Alls & 5-Qubit & \\
    of & per Disk & per Disk & Kernels per & Run Time \\
    Cycles & Slice & Slice & Disk Slice & (days) \\
    \hline
    10 &  1 &  3.002 &  65 & 0.67 \\
    14 &  3 &  6.002 &  89 & 1.61 \\
    20 &  5 &  9.002 & 120 & 2.55 \\
    24 &  7 & 13.002 & 141 & 3.54 \\
    28 &  9 & 16.002 & 162 & 4.47 \\
    32 & 11 & 20.002 & 182 & 5.46 \\
    36 & 13 & 24.002 & 206 & 6.45 \\
    \hline
  \end{tabular}
  \caption{Estimates of total run times for simulating 53-qubit,
    Sycamore ABCDCDAB circuits of various depths.}
  \label{tab:53_manycycles}
\end{table}

\begin{table}[tbp]
  \centering
  \begin{tabular}{|c|c|c|c|c|}
    \hline
    Number & Disk Xfers & All-to-Alls & 5-Qubit & \\
    of & per Disk & per Disk & Kernels per & Run Time \\
    Cycles & Slice & Slice & Disk Slice & (days) \\
    \hline
    10 &  1 &  3.004 &  66 &  2.05 \\
    14 &  3 &  6.004 &  90 &  3.92 \\
    20 &  5 &  9.004 & 122 &  5.80 \\
    24 &  7 & 13.004 & 144 &  7.78 \\
    28 &  9 & 16.004 & 166 &  9.65 \\
    32 & 11 & 20.004 & 187 & 11.63 \\
    36 & 13 & 24.004 & 211 & 13.62 \\
    \hline
  \end{tabular}
  \caption{Estimates of total run times for simulating 54-qubit,
    Sycamore ABCDCDAB circuits of various depths.}
  \label{tab:54_manycycles}
\end{table}

\afterpage{\clearpage}



\bibliography{breaking_54qubit}

\bibliographystyle{abbrv}

\clearpage

\appendix

\section{Implementation details}
We give a full description of the implementation of the tensor
computations, including cache blocking to create the 5-qubit aggregate
gates. Further optimizations might be possible; this description is
meant as a viable proof of concept modeled on
\cite{haner2017simulation}.

We describe how to interpret each line in the listing. Qubits are
numbered from $0$ to $n-1$. The ``Mode'' column indicates the type of
information contained in the corresponding line. We give an overview
of each of the eight possible modes.
\begin{itemize}
\item ``define'': the first define line indicates the total number of
  qubits, the second define line describes the qubit indices used to
  address logical files on disk.
\item ``new'':
  \begin{itemize}
  \item if the ``Gate'' column contains ``tensor'', it describes a new
    tensor. The column ``Arguments'' indicates the corresponding
    number of local qubits and of global qubits, followed by the
    indices of the qubits in the tensor, starting with local qubits
    and ending with global qubits.
  \item if the ``Gate'' column contains ``cache'', it indicates how to
    partition gates into 5-qubit aggregate gates.
  \end{itemize}
\item ``gate'': describes a gate. All gates are two-qubit gates after
  circuit transformations, as discussed in the paper; the ``Arguments''
  column indicates the qubits involved.
\item ``entgl'': describes the introduction of an entanglement index
  due to a deferred contraction between tensors.
  \begin{itemize}
  \item if the ``Gate'' column contains ``tensor'', it describes the
    new tensor with the corresponding list of entanglement indices
    (labeled with negative numbers).
  \item if the ``Gate'' column contains ``EI'' or ``E2Q'', it
    describes synthesized gate operations that are employed when
    introducing entanglement indices.
  \end{itemize}
\item ``slice'': lists the indices of the qubits that are sliced on
  the first level of the recursive scheme.
\item ``all2all'': indicates a communication between nodes used to
  rearrange tensor indices in preparation for a contraction, or to
  swap which qubits are local and which ones are global.
\item ``write'': indicates which qubit indices to fix, in order to
  write a slice to disk.
\item ``read'': indicates which qubit indices to fix, in order to
  read a slice from disk.
\end{itemize}

Note that in the listings, gate $2Q$ together with the level of the
gate in the circuit and the qubits to which that gate is applied
refers to a specific instance of one of the gates shown in
Figs.~\ref{fig:53_gatepattern} and~\ref{fig:54_gatepattern} with its
own potentially unique associated unitary matrix.

To define the synthesized gate operations $EI$ and $E2Q$ , suppose one
of these $2Q$ gate bridges qubits $a$ and $b$ in Tensors 1 and 2,
respectively.  Let $\phi_a$ and $\chi_b$ be the corresponding tensors
prior to applying that $2Q$ gate.  Then the resulting quantum state is
given by
\begin{equation}
  \psi_{a^{\prime\prime},b^{\prime\prime}} = \sum_{a,b} 2Q_{a^{\prime\prime},b^{\prime\prime},a,b}\cdot\phi_a\cdot\chi_b
  = \sum_{a^\prime a,b} I_{a^{\prime\prime},a^\prime}\cdot 2Q_{a^\prime,b^{\prime\prime},a,b}\cdot\phi_a\cdot \chi_b
\end{equation}
where $I$ is the identity matrix. The above equation can be rewritten as
\begin{equation}
  \psi_{a^{\prime\prime},b^{\prime\prime}} = \sum_{a^{\prime},a} \phi^{\prime}_{a^{\prime\prime},a^{\prime},a}
  \cdot \chi^{\prime}_{b^{\prime\prime},a^{\prime},a}
\end{equation}
where
\begin{equation}
  \phi^{\prime}_{a^{\prime\prime},a^{\prime},a} = I_{a^{\prime\prime},a^\prime}\cdot\phi_a \ ,\qquad
  \chi^{\prime}_{b^{\prime\prime},a^{\prime},a} =  \sum_{b} {2Q_{a^{\prime},b^{\prime\prime},a,b}}\cdot\chi_b
\end{equation}
These last two equations define the $EI$ and $E2Q$ synthesized gate
operations, respectively, where $a''$ is effectively the new index for
qubit $a$, $b''$ the new index for qubit $b$, and $a'$ and $a$ are
entanglement indices introduced through these synthesized gate
operations.  The equation above it defines the contraction performed
to eliminate entanglement indices.

\vskip2ex

The following listing is for the 53-qubit circuit.
{\scriptsize
\begin{longtable}{|c|c|c|c|c|l|}
  \hline
  \hline
  Tensor&Phase&Mode&Depth&Gate&Arguments\\
  \hline
  \endfirsthead
  \hline
  Tensor&Phase&Mode&Depth&Gate&Arguments\\
  \hline
  \endhead
  \hline
  \endfoot
  \hline
  \hline
  \endlastfoot
  
  0&0&define&0&qubits&53\\
  0&0&define&0&disk&0,1,2,3,4,5,6,7,8,23,24,25,26,27,28\\
  &&&&&\quad ,29,30,49,50,51,52\\
  1&1&new&0&tensor&27,0,0,1,2,3,4,5,6,7,8,9,10,11,12\\
  &&&&&\quad ,13,14,15,16,17,18,19,20,21,22,23,24,25,26\\
  1&1&new&0&cache&10,14,18,19,23\\
  1&1&gate&1&2Q&14,10\\
  1&1&gate&1&2Q&23,19\\
  1&1&gate&3&2Q&18,14\\
  1&1&gate&6&2Q&23,18\\
  1&1&new&0&cache&2,6,10,18,23\\
  1&1&gate&2&2Q&6,2\\
  1&1&gate&4&2Q&10,6\\
  1&1&new&0&cache&1,5,6,18,23\\
  1&1&gate&2&2Q&5,1\\
  1&1&gate&6&2Q&6,1\\
  1&1&new&0&cache&5,9,14,18,23\\
  1&1&gate&3&2Q&9,5\\
  1&1&gate&6&2Q&14,9\\
  1&1&new&0&cache&0,5,18,23\\
  1&1&gate&5&2Q&5,0\\
  1&1&new&0&cache&18,22,23,26\\
  1&1&gate&2&2Q&26,22\\
  1&1&new&0&cache&18,20,23,24,26\\
  1&1&gate&1&2Q&24,20\\
  1&1&new&0&cache&12,16,18,20,23\\
  1&1&gate&2&2Q&16,12\\
  1&1&gate&4&2Q&20,16\\
  1&1&new&0&cache&18,20,21,23,25\\
  1&1&gate&2&2Q&25,21\\
  1&2&entgl&0&tensor&27,8,0,1,2,3,4,5,6,7,8,9,10,11,12\\
  &&&&&\quad ,13,14,15,16,17,18,19,20,21,22,23,24,25,26,-1\\
  &&&&&\quad ,-2,-3,-4,-5,-6,-7,-8,-9,-10,-11,-12,-13,-14\\
  1&2&entgl&3&EI&25,-1,-2\\
  1&2&entgl&3&EI&24,-3,-4\\
  1&2&entgl&4&EI&26,-5,-6\\
  1&2&entgl&7&EI&23,-9,-10\\
  1&2&new&0&cache&18,20,23,25\\
  1&2&gate&5&2Q&25,20\\
  1&2&gate&10&2Q&23,18\\
  1&2&new&0&cache&11,14,15,19,24\\
  1&2&gate&2&2Q&15,11\\
  1&2&gate&3&2Q&19,15\\
  1&2&gate&5&2Q&24,19\\
  1&2&gate&8&2Q&19,14\\
  1&2&new&0&cache&9,14,19,24,25\\
  1&2&gate&10&2Q&14,9\\
  1&3&entgl&0&tensor&27,12,0,1,2,3,4,5,6,7,8,9,10,11,12\\
  &&&&&\quad ,13,14,15,16,17,18,19,20,21,22,23,24,25,26,-1\\
  &&&&&\quad ,-2,-3,-4,-5,-6,-7,-8,-9,-10,-11,-12,-13,-14\\
  1&3&entgl&7&EI&24,-7,-8\\
  1&3&entgl&8&EI&25,-13,-14\\
  1&3&new&0&cache&10,14,15,19,24\\
  1&3&gate&6&2Q&15,10\\
  1&3&gate&9&2Q&24,19\\
  1&3&gate&12&2Q&19,14\\
  1&3&new&0&cache&0,5,10,15,20\\
  1&3&gate&7&2Q&20,15\\
  1&3&gate&8&2Q&10,5\\
  1&3&gate&9&2Q&5,0\\
  1&3&gate&10&2Q&15,10\\
  1&3&gate&12&2Q&10,5\\
  1&3&new&0&cache&10,14,18,20,25\\
  1&3&gate&9&2Q&25,20\\
  1&3&gate&13&2Q&14,10\\
  1&3&gate&15&2Q&18,14\\
  1&3&new&0&cache&13,15,17,20,21\\
  1&3&gate&1&2Q&17,13\\
  1&3&gate&4&2Q&21,17\\
  1&3&gate&11&2Q&20,15\\
  1&3&new&0&cache&3,7,11,21,26\\
  1&3&gate&1&2Q&7,3\\
  1&3&gate&4&2Q&11,7\\
  1&3&gate&6&2Q&26,21\\
  1&3&new&0&cache&1,5,6,11,16\\
  1&3&gate&5&2Q&16,11\\
  1&3&gate&8&2Q&11,6\\
  1&3&gate&10&2Q&6,1\\
  1&3&gate&14&2Q&5,1\\
  1&3&new&0&cache&0,1,5,9,26\\
  1&3&gate&15&2Q&9,5\\
  1&3&gate&18&2Q&5,1\\
  1&3&gate&19&2Q&9,5\\
  1&3&gate&21&2Q&5,0\\
  1&3&new&0&cache&6,11,15,16,21\\
  1&3&gate&7&2Q&21,16\\
  1&3&gate&9&2Q&16,11\\
  1&3&gate&12&2Q&11,6\\
  1&3&gate&14&2Q&15,11\\
  1&3&new&0&cache&3,4,8,12,13\\
  1&3&gate&1&2Q&8,4\\
  1&3&gate&3&2Q&12,8\\
  1&3&gate&5&2Q&8,3\\
  1&3&gate&7&2Q&13,8\\
  1&3&gate&9&2Q&8,3\\
  1&3&gate&11&2Q&13,8\\
  1&3&gate&13&2Q&8,4\\
  1&3&new&0&cache&2,6,7,12,17\\
  1&3&gate&5&2Q&17,12\\
  1&3&gate&6&2Q&7,2\\
  1&3&gate&7&2Q&12,7\\
  1&3&gate&10&2Q&7,2\\
  1&3&gate&14&2Q&6,2\\
  1&3&new&0&cache&6,9,10,14,18\\
  1&3&gate&16&2Q&10,6\\
  1&3&gate&17&2Q&14,10\\
  1&3&gate&19&2Q&18,14\\
  1&3&gate&22&2Q&14,9\\
  1&3&new&0&cache&1,2,6,10,26\\
  1&3&gate&18&2Q&6,2\\
  1&3&gate&20&2Q&10,6\\
  1&3&gate&22&2Q&6,1\\
  1&3&new&0&cache&7,12,13,17,22\\
  1&3&gate&8&2Q&22,17\\
  1&3&gate&9&2Q&17,12\\
  1&3&gate&11&2Q&12,7\\
  1&3&gate&12&2Q&22,17\\
  1&3&gate&13&2Q&17,13\\
  1&3&new&0&cache&3,7,11,21,26\\
  1&3&gate&13&2Q&7,3\\
  1&3&gate&16&2Q&11,7\\
  1&3&gate&17&2Q&7,3\\
  1&4&entgl&0&tensor&28,13,0,1,2,3,4,5,6,7,8,9,10,11,12\\
  &&&&&\quad ,13,14,15,16,17,18,19,20,21,22,23,24,25,26,-1\\
  &&&&&\quad ,-2,-3,-4,-5,-6,-7,-8,-9,-10,-11,-12,-13,-14\\
  1&4&entgl&8&EI&26,-11,-12\\
  1&4&new&0&cache&8,12,16,21,26\\
  1&4&gate&10&2Q&26,21\\
  1&4&gate&11&2Q&21,16\\
  1&4&gate&14&2Q&16,12\\
  1&4&gate&15&2Q&12,8\\
  1&4&new&0&cache&4,8\\
  1&4&gate&17&2Q&8,4\\
  2&5&new&0&tensor&26,0,27,28,29,30,31,32,33,34,35,36,37,38,39\\
  &&&&&\quad ,40,41,42,43,44,45,46,47,48,49,50,51,52\\
  2&5&new&0&cache&36,40,44,45,49\\
  2&5&gate&1&2Q&49,45\\
  2&5&gate&2&2Q&40,36\\
  2&5&gate&3&2Q&44,40\\
  2&5&gate&6&2Q&49,44\\
  2&5&gate&10&2Q&49,44\\
  2&5&new&0&cache&30,34,37,41,45\\
  2&5&gate&2&2Q&34,30\\
  2&5&gate&2&2Q&41,37\\
  2&5&gate&4&2Q&45,41\\
  2&5&new&0&cache&29,33,34,38,42\\
  2&5&gate&1&2Q&33,29\\
  2&5&gate&1&2Q&42,38\\
  2&5&gate&3&2Q&38,34\\
  2&5&new&0&cache&29,30,34,39,43\\
  2&5&gate&1&2Q&43,39\\
  2&6&entgl&0&tensor&26,4,27,28,29,30,31,32,33,34,35,36,37,38,39\\
  &&&&&\quad ,40,41,42,43,44,45,46,47,48,49,50,51,52,-1,-2\\
  &&&&&\quad ,-3,-4,-5,-6,-7,-8,-9,-10,-11,-12,-13,-14\\
  2&6&entgl&4&E2Q&29,-5,-6\\
  2&6&entgl&8&E2Q&30,-11,-12\\
  2&6&new&0&cache&29,33,34,37,39\\
  2&6&gate&3&2Q&37,33\\
  2&6&gate&6&2Q&34,29\\
  2&6&gate&7&2Q&39,34\\
  2&6&new&0&cache&27,29,31,34,39\\
  2&6&gate&2&2Q&31,27\\
  2&6&new&0&cache&29,31,34,35,39\\
  2&6&gate&4&2Q&35,31\\
  2&6&new&0&cache&35,40,45,46,50\\
  2&6&gate&2&2Q&50,46\\
  2&6&gate&5&2Q&40,35\\
  2&6&gate&6&2Q&50,45\\
  2&6&gate&8&2Q&45,40\\
  2&6&gate&9&2Q&40,35\\
  2&6&gate&10&2Q&50,45\\
  2&6&gate&12&2Q&45,40\\
  2&6&new&0&cache&29,34,39,45,49\\
  2&6&gate&13&2Q&49,45\\
  2&6&new&0&cache&29,34,39,42,46\\
  2&6&gate&4&2Q&46,42\\
  2&6&new&0&cache&29,34,37,39,42\\
  2&6&gate&6&2Q&42,37\\
  2&7&entgl&0&tensor&26,8,27,28,29,30,31,32,33,34,35,36,37,38,39\\
  &&&&&\quad ,40,41,42,43,44,45,46,47,48,49,50,51,52,-1,-2\\
  &&&&&\quad ,-3,-4,-5,-6,-7,-8,-9,-10,-11,-12,-13,-14\\
  2&7&entgl&3&E2Q&27,-3,-4\\
  2&7&entgl&8&E2Q&29,-13,-14\\
  2&7&new&0&cache&28,29,32,34,39\\
  2&7&gate&1&2Q&32,28\\
  2&7&gate&10&2Q&34,29\\
  2&7&gate&11&2Q&39,34\\
  2&7&new&0&cache&27,32,36,37,41\\
  2&7&gate&4&2Q&36,32\\
  2&7&gate&5&2Q&32,27\\
  2&7&gate&6&2Q&41,36\\
  2&7&gate&8&2Q&37,32\\
  2&7&new&0&cache&27,41,46,47,51\\
  2&7&gate&2&2Q&51,47\\
  2&7&gate&5&2Q&51,46\\
  2&7&gate&8&2Q&46,41\\
  2&7&gate&9&2Q&51,46\\
  2&7&new&0&cache&31,36,40,41,44\\
  2&7&gate&7&2Q&36,31\\
  2&7&gate&10&2Q&41,36\\
  2&7&gate&11&2Q&36,31\\
  2&7&gate&14&2Q&40,36\\
  2&7&gate&15&2Q&44,40\\
  2&7&new&0&cache&27,32,41,46,50\\
  2&7&gate&12&2Q&46,41\\
  2&7&gate&14&2Q&50,46\\
  2&7&new&0&cache&28,33,38,43,47\\
  2&7&gate&3&2Q&47,43\\
  2&7&gate&5&2Q&43,38\\
  2&7&new&0&cache&37,42,47,48,52\\
  2&7&gate&1&2Q&52,48\\
  2&7&gate&5&2Q&52,47\\
  2&7&gate&7&2Q&47,42\\
  2&7&gate&9&2Q&52,47\\
  2&7&gate&10&2Q&42,37\\
  2&7&gate&11&2Q&47,42\\
  2&8&entgl&0&tensor&26,12,27,28,29,30,31,32,33,34,35,36,37,38,39\\
  &&&&&\quad ,40,41,42,43,44,45,46,47,48,49,50,51,52,-1,-2\\
  &&&&&\quad ,-3,-4,-5,-6,-7,-8,-9,-10,-11,-12,-13,-14\\
  2&8&entgl&3&E2Q&28,-1,-2\\
  2&8&entgl&7&E2Q&27,-9,-10\\
  2&8&new&0&cache&27,32,37,41,45\\
  2&8&gate&9&2Q&32,27\\
  2&8&gate&12&2Q&37,32\\
  2&8&gate&14&2Q&41,37\\
  2&8&gate&16&2Q&45,41\\
  2&8&new&0&cache&45,47,49,51\\
  2&8&gate&14&2Q&51,47\\
  2&8&gate&17&2Q&49,45\\
  2&8&new&0&cache&28,33,38,43,48\\
  2&8&gate&6&2Q&33,28\\
  2&8&gate&7&2Q&48,43\\
  2&8&gate&8&2Q&38,33\\
  2&8&gate&9&2Q&43,38\\
  2&8&gate&11&2Q&48,43\\
  2&8&new&0&cache&28,39,43,47,51\\
  2&8&gate&13&2Q&43,39\\
  2&8&gate&15&2Q&47,43\\
  2&8&gate&17&2Q&43,39\\
  2&8&gate&18&2Q&51,47\\
  2&8&gate&19&2Q&47,43\\
  2&8&new&0&cache&28,33,47,48,52\\
  2&8&gate&13&2Q&52,48\\
  2&8&gate&17&2Q&52,48\\
  2&8&gate&21&2Q&52,47\\
  2&9&entgl&0&tensor&27,13,27,28,29,30,31,32,33,34,35,36,37,38,39\\
  &&&&&\quad ,40,41,42,43,44,45,46,47,48,49,50,51,52,-1,-2\\
  &&&&&\quad ,-3,-4,-5,-6,-7,-8,-9,-10,-11,-12,-13,-14\\
  2&9&entgl&7&E2Q&28,-7,-8\\
  2&9&new&0&cache&28,33,38,42,46\\
  2&9&gate&10&2Q&33,28\\
  2&9&gate&12&2Q&38,33\\
  2&9&gate&13&2Q&42,38\\
  2&9&gate&16&2Q&46,42\\
  2&9&new&0&cache&46,50\\
  2&9&gate&18&2Q&50,46\\
  0&10&slice&0&disk&0,1,2,3,49,50,51,52\\
  1&10&all2all&0&tensor&30,7,-1,-2,-3,-4,-5,-6,-7,-8,-9,-10,-11,-12,-13\\
  &&&&&\quad ,-14,11,12,13,14,15,16,17,18,19,20,21,22,23,24\\
  &&&&&\quad ,25,26,4,5,6,7,8,9,10\\
  2&10&all2all&0&tensor&30,6,-1,-2,-3,-4,-5,-6,-7,-8,-9,-10,-11,-12,-13\\
  &&&&&\quad ,-14,27,28,29,30,31,32,33,34,35,36,37,38,39,40\\
  &&&&&\quad ,41,42,43,44,45,46,47,48\\
  3&1&new&0&tensor&32,13,11,12,13,14,15,16,17,18,19,20,21,22,23\\
  &&&&&\quad ,24,25,26,27,28,29,30,31,32,33,34,35,36,37,38\\
  &&&&&\quad ,39,40,41,42,4,5,6,7,8,9,10,43,44,45,46\\
  &&&&&\quad ,47,48\\
  3&1&new&0&cache&15,18,19,23,27\\
  3&1&gate&11&2Q&27,23\\
  3&1&gate&13&2Q&23,19\\
  3&1&gate&15&2Q&19,15\\
  3&1&gate&17&2Q&23,19\\
  3&1&gate&22&2Q&23,18\\
  3&1&new&0&cache&26,30,34,38,42\\
  3&1&gate&12&2Q&30,26\\
  3&1&gate&14&2Q&34,30\\
  3&1&gate&15&2Q&38,34\\
  3&1&gate&17&2Q&42,38\\
  3&1&gate&18&2Q&34,30\\
  3&1&gate&19&2Q&38,34\\
  3&1&new&0&cache&11,15,19,27,31\\
  3&1&gate&14&2Q&31,27\\
  3&1&gate&18&2Q&15,11\\
  3&1&gate&19&2Q&19,15\\
  3&1&new&0&cache&24,28,32,36,40\\
  3&1&gate&11&2Q&28,24\\
  3&1&gate&13&2Q&32,28\\
  3&1&gate&16&2Q&36,32\\
  3&1&gate&18&2Q&40,36\\
  3&1&new&0&cache&20,24,27,31,35\\
  3&1&gate&13&2Q&24,20\\
  3&1&gate&15&2Q&27,24\\
  3&1&gate&16&2Q&35,31\\
  3&1&gate&18&2Q&31,27\\
  3&1&gate&20&2Q&35,31\\
  3&1&new&0&cache&16,19,20,24,27\\
  3&1&gate&16&2Q&20,16\\
  3&1&gate&17&2Q&24,20\\
  3&1&gate&19&2Q&27,24\\
  3&1&gate&21&2Q&24,19\\
  3&1&new&0&cache&12,14,16,19,20\\
  3&1&gate&18&2Q&16,12\\
  3&1&gate&20&2Q&20,16\\
  3&1&gate&24&2Q&19,14\\
  3&1&new&0&cache&25,29,33,37,41\\
  3&1&gate&12&2Q&29,25\\
  3&1&gate&13&2Q&33,29\\
  3&1&gate&15&2Q&37,33\\
  3&1&gate&18&2Q&41,37\\
  3&1&new&0&cache&22,26,29,33,34\\
  3&1&gate&14&2Q&26,22\\
  3&1&gate&16&2Q&29,26\\
  3&1&gate&17&2Q&33,29\\
  3&1&gate&18&2Q&26,22\\
  3&1&gate&20&2Q&29,26\\
  3&1&gate&22&2Q&34,29\\
  3&1&new&0&cache&33,34,37,39\\
  3&1&gate&19&2Q&37,33\\
  3&1&gate&23&2Q&39,34\\
  3&1&new&0&cache&21,25,28,32,36\\
  3&1&gate&14&2Q&25,21\\
  3&1&gate&15&2Q&28,25\\
  3&1&gate&17&2Q&32,28\\
  3&1&gate&20&2Q&36,32\\
  3&1&new&0&cache&18,23,27,32\\
  3&1&gate&21&2Q&32,27\\
  3&1&gate&23&2Q&27,23\\
  3&1&gate&26&2Q&23,18\\
  3&1&new&0&cache&17,21,25,28,33\\
  3&1&gate&16&2Q&21,17\\
  3&1&gate&18&2Q&25,21\\
  3&1&gate&19&2Q&28,25\\
  3&1&gate&22&2Q&33,28\\
  3&1&new&0&cache&20,25,29,34,39\\
  3&1&gate&21&2Q&25,20\\
  3&1&gate&24&2Q&29,25\\
  3&1&gate&26&2Q&34,29\\
  3&1&gate&27&2Q&39,34\\
  3&1&new&0&cache&13,17,19,24,28\\
  3&1&gate&17&2Q&17,13\\
  3&1&gate&23&2Q&28,24\\
  3&1&gate&25&2Q&24,19\\
  3&1&new&0&cache&17,21,26,30\\
  3&1&gate&20&2Q&21,17\\
  3&1&gate&22&2Q&26,21\\
  3&1&gate&24&2Q&30,26\\
  3&2&all2all&0&tensor&32,13,17,18,19,20,21,22,23,24,25,26,27,28,29\\
  &&&&&\quad ,30,31,32,33,34,35,36,37,38,39,40,41,42,43,44\\
  &&&&&\quad ,45,46,47,48,4,5,6,7,8,9,10,11,12,13,14\\
  &&&&&\quad ,15,16\\
  3&2&new&0&cache&33,38,39,43,48\\
  3&2&gate&21&2Q&43,38\\
  3&2&gate&23&2Q&48,43\\
  3&2&gate&24&2Q&38,33\\
  3&2&gate&25&2Q&43,38\\
  3&2&gate&27&2Q&48,43\\
  3&2&gate&29&2Q&43,39\\
  3&2&new&0&cache&24,28,33,38\\
  3&2&gate&26&2Q&33,28\\
  3&2&gate&27&2Q&28,24\\
  3&2&gate&28&2Q&38,33\\
  3&2&new&0&cache&32,37,42,46,47\\
  3&2&gate&20&2Q&46,42\\
  3&2&gate&22&2Q&42,37\\
  3&2&gate&23&2Q&47,42\\
  3&2&gate&24&2Q&37,32\\
  3&2&gate&26&2Q&42,37\\
  3&2&new&0&cache&23,27,28,32,37\\
  3&2&gate&25&2Q&32,27\\
  3&2&gate&27&2Q&27,23\\
  3&2&gate&28&2Q&37,32\\
  3&2&gate&29&2Q&32,28\\
  3&2&new&0&cache&31,36,41,45\\
  3&2&gate&20&2Q&45,41\\
  3&2&gate&22&2Q&41,36\\
  3&2&gate&23&2Q&36,31\\
  3&2&new&0&cache&35,40,44\\
  3&2&gate&19&2Q&44,40\\
  3&2&gate&21&2Q&40,35\\
  3&3&all2all&0&tensor&32,13,4,5,6,7,8,9,10,11,12,13,14,15,16\\
  &&&&&\quad ,17,18,19,20,21,22,23,24,25,26,27,28,29,30,31\\
  &&&&&\quad ,32,33,34,35,36,37,38,39,40,41,42,43,44,45,46\\
  &&&&&\quad ,47,48\\
  3&3&new&0&cache&9,14,19,23\\
  3&3&gate&26&2Q&14,9\\
  3&3&gate&28&2Q&19,14\\
  3&3&gate&29&2Q&23,19\\
  3&3&new&0&cache&10,15,20,25,29\\
  3&3&gate&22&2Q&15,10\\
  3&3&gate&23&2Q&20,15\\
  3&3&gate&25&2Q&25,20\\
  3&3&gate&28&2Q&29,25\\
  3&3&new&0&cache&5,10,15,29,33\\
  3&3&gate&24&2Q&10,5\\
  3&3&gate&26&2Q&15,10\\
  3&3&gate&29&2Q&33,29\\
  3&3&new&0&cache&7,11,15,20,24\\
  3&3&gate&20&2Q&11,7\\
  3&3&gate&27&2Q&20,15\\
  3&3&gate&29&2Q&24,20\\
  3&3&new&0&cache&11,16,21,26,30\\
  3&3&gate&21&2Q&16,11\\
  3&3&gate&23&2Q&21,16\\
  3&3&gate&26&2Q&26,21\\
  3&3&gate&28&2Q&30,26\\
  3&3&new&0&cache&6,11,16,30,34\\
  3&3&gate&24&2Q&11,6\\
  3&3&gate&25&2Q&16,11\\
  3&3&gate&30&2Q&34,30\\
  3&3&new&0&cache&16,21,25,28\\
  3&3&gate&27&2Q&21,16\\
  3&3&gate&30&2Q&25,21\\
  3&3&gate&31&2Q&28,25\\
  3&3&new&0&cache&8,12,17,22\\
  3&3&gate&19&2Q&12,8\\
  3&3&gate&21&2Q&17,12\\
  3&3&gate&24&2Q&22,17\\
  3&3&write&0&disk&0,1,2,3,49,50,51,52\\
  4&1&read&0&disk&23,24,25,26,27,28,29,30\\
  4&1&new&0&tensor&32,13,0,1,2,3,4,5,6,7,8,9,10,11,12\\
  &&&&&\quad ,13,14,15,16,17,18,19,20,21,22,31,32,33,34,35\\
  &&&&&\quad ,36,37,38,39,40,41,42,43,44,45,46,47,48,49,50\\
  &&&&&\quad ,51,52\\
  4&1&new&0&cache&0,5,10,14,18\\
  4&1&gate&25&2Q&5,0\\
  4&1&gate&28&2Q&10,5\\
  4&1&gate&29&2Q&14,10\\
  4&1&gate&31&2Q&18,14\\
  4&1&new&0&cache&0,1,5,6,9\\
  4&1&gate&26&2Q&6,1\\
  4&1&gate&30&2Q&5,1\\
  4&1&gate&31&2Q&9,5\\
  4&1&gate&34&2Q&5,1\\
  4&1&gate&35&2Q&9,5\\
  4&1&gate&37&2Q&5,0\\
  4&1&new&0&cache&6,11,15,19\\
  4&1&gate&28&2Q&11,6\\
  4&1&gate&30&2Q&15,11\\
  4&1&gate&31&2Q&19,15\\
  4&1&new&0&cache&2,6,7,10,12\\
  4&1&gate&22&2Q&7,2\\
  4&1&gate&23&2Q&12,7\\
  4&1&gate&26&2Q&7,2\\
  4&1&gate&30&2Q&6,2\\
  4&1&gate&32&2Q&10,6\\
  4&1&gate&34&2Q&6,2\\
  4&1&new&0&cache&6,9,10,14,18\\
  4&1&gate&33&2Q&14,10\\
  4&1&gate&35&2Q&18,14\\
  4&1&gate&36&2Q&10,6\\
  4&1&gate&38&2Q&14,9\\
  4&1&new&0&cache&1,6,12,17,22\\
  4&1&gate&25&2Q&17,12\\
  4&1&gate&28&2Q&22,17\\
  4&1&gate&38&2Q&6,1\\
  4&1&new&0&cache&3,8,13,17,21\\
  4&1&gate&21&2Q&8,3\\
  4&1&gate&23&2Q&13,8\\
  4&1&gate&25&2Q&8,3\\
  4&1&gate&27&2Q&13,8\\
  4&1&gate&29&2Q&17,13\\
  4&1&gate&32&2Q&21,17\\
  4&1&gate&33&2Q&17,13\\
  4&1&new&0&cache&3,7,11,12,15\\
  4&1&gate&27&2Q&12,7\\
  4&1&gate&29&2Q&7,3\\
  4&1&gate&32&2Q&11,7\\
  4&1&gate&33&2Q&7,3\\
  4&1&gate&34&2Q&15,11\\
  4&1&gate&36&2Q&11,7\\
  4&1&new&0&cache&2,7,12,16,20\\
  4&1&gate&30&2Q&16,12\\
  4&1&gate&32&2Q&20,16\\
  4&1&gate&38&2Q&7,2\\
  4&1&new&0&cache&3,4,8,12,16\\
  4&1&gate&29&2Q&8,4\\
  4&1&gate&31&2Q&12,8\\
  4&1&gate&33&2Q&8,4\\
  4&1&gate&34&2Q&16,12\\
  4&1&gate&35&2Q&12,8\\
  4&1&gate&37&2Q&8,3\\
  4&1&new&0&cache&8,13\\
  4&1&gate&39&2Q&13,8\\
  4&2&all2all&0&tensor&32,13,13,14,15,16,17,18,19,20,21,22,31,32,33\\
  &&&&&\quad ,34,35,36,37,38,39,40,41,42,43,44,45,46,47,48\\
  &&&&&\quad ,49,50,51,52,0,1,2,3,4,5,6,7,8,9,10\\
  &&&&&\quad ,11,12\\
  4&2&new&0&cache&38,42,47,48,52\\
  4&2&gate&25&2Q&52,47\\
  4&2&gate&27&2Q&47,42\\
  4&2&gate&29&2Q&42,38\\
  4&2&gate&29&2Q&52,48\\
  4&2&gate&33&2Q&52,48\\
  4&2&new&0&cache&41,43,46,47,51\\
  4&2&gate&21&2Q&51,46\\
  4&2&gate&24&2Q&46,41\\
  4&2&gate&25&2Q&51,46\\
  4&2&gate&30&2Q&51,47\\
  4&2&gate&31&2Q&47,43\\
  4&2&gate&34&2Q&51,47\\
  4&2&new&0&cache&39,43,47,52\\
  4&2&gate&33&2Q&43,39\\
  4&2&gate&35&2Q&47,43\\
  4&2&gate&37&2Q&52,47\\
  4&2&new&0&cache&33,36,37,41,46\\
  4&2&gate&26&2Q&41,36\\
  4&2&gate&28&2Q&46,41\\
  4&2&gate&30&2Q&41,37\\
  4&2&gate&31&2Q&37,33\\
  4&2&new&0&cache&40,42,45,46,50\\
  4&2&gate&22&2Q&50,45\\
  4&2&gate&24&2Q&45,40\\
  4&2&gate&26&2Q&50,45\\
  4&2&gate&30&2Q&50,46\\
  4&2&gate&32&2Q&46,42\\
  4&2&gate&34&2Q&50,46\\
  4&2&new&0&cache&34,38,42,46,51\\
  4&2&gate&31&2Q&38,34\\
  4&2&gate&33&2Q&42,38\\
  4&2&gate&36&2Q&46,42\\
  4&2&gate&37&2Q&51,46\\
  4&2&new&0&cache&31,35,36,40,45\\
  4&2&gate&25&2Q&40,35\\
  4&2&gate&27&2Q&36,31\\
  4&2&gate&28&2Q&45,40\\
  4&2&gate&30&2Q&40,36\\
  4&2&new&0&cache&37,41,44,45,49\\
  4&2&gate&22&2Q&49,44\\
  4&2&gate&26&2Q&49,44\\
  4&2&gate&29&2Q&49,45\\
  4&2&gate&32&2Q&45,41\\
  4&2&gate&33&2Q&49,45\\
  4&2&gate&34&2Q&41,37\\
  4&2&gate&36&2Q&45,41\\
  4&2&new&0&cache&32,36,40,44,49\\
  4&2&gate&31&2Q&44,40\\
  4&2&gate&32&2Q&36,32\\
  4&2&gate&34&2Q&40,36\\
  4&2&gate&35&2Q&44,40\\
  4&2&gate&38&2Q&49,44\\
  4&2&new&0&cache&45,50\\
  4&2&gate&38&2Q&50,45\\
  4&2&write&0&disk&23,24,25,26,27,28,29,30\\
  5&1&read&0&disk&0,1,2,3,49,50,51,52\\
  5&1&new&0&tensor&32,13,4,5,6,7,8,9,10,11,12,13,14,15,16\\
  &&&&&\quad ,17,18,19,20,21,22,23,24,25,26,27,28,29,30,31\\
  &&&&&\quad ,32,33,34,35,36,37,38,39,40,41,42,43,44,45,46\\
  &&&&&\quad ,47,48\\
  5&1&new&0&cache&5,10,15,19,23\\
  5&1&gate&33&2Q&23,19\\
  5&1&gate&35&2Q&19,15\\
  5&1&gate&38&2Q&15,10\\
  5&1&gate&40&2Q&10,5\\
  5&1&new&0&cache&7,12,17,21,25\\
  5&1&gate&34&2Q&25,21\\
  5&1&gate&36&2Q&21,17\\
  5&1&gate&37&2Q&17,12\\
  5&1&gate&39&2Q&12,7\\
  5&1&new&0&cache&17,22,26,29,33\\
  5&1&gate&30&2Q&26,22\\
  5&1&gate&32&2Q&29,26\\
  5&1&gate&33&2Q&33,29\\
  5&1&gate&34&2Q&26,22\\
  5&1&gate&36&2Q&29,26\\
  5&1&gate&40&2Q&22,17\\
  5&1&new&0&cache&21,26,30,34\\
  5&1&gate&34&2Q&34,30\\
  5&1&gate&38&2Q&26,21\\
  5&1&gate&40&2Q&30,26\\
  5&1&new&0&cache&18,23,25,28,32\\
  5&1&gate&33&2Q&32,28\\
  5&1&gate&35&2Q&28,25\\
  5&1&gate&38&2Q&23,18\\
  5&1&new&0&cache&20,24,27,31,35\\
  5&1&gate&30&2Q&31,27\\
  5&1&gate&31&2Q&27,24\\
  5&1&gate&32&2Q&35,31\\
  5&1&gate&33&2Q&24,20\\
  5&1&gate&34&2Q&31,27\\
  5&1&gate&35&2Q&27,24\\
  5&1&gate&36&2Q&35,31\\
  5&1&new&0&cache&14,16,19,20,24\\
  5&1&gate&36&2Q&20,16\\
  5&1&gate&37&2Q&24,19\\
  5&1&gate&40&2Q&19,14\\
  5&1&new&0&cache&6,11,16,21\\
  5&1&gate&37&2Q&16,11\\
  5&1&gate&39&2Q&21,16\\
  5&1&gate&40&2Q&11,6\\
  5&1&new&0&cache&15,20,25\\
  5&1&gate&37&2Q&25,20\\
  5&1&gate&39&2Q&20,15\\
  5&2&all2all&0&tensor&32,13,17,18,19,20,21,22,23,24,25,26,27,28,29\\
  &&&&&\quad ,30,31,32,33,34,35,36,37,38,39,40,41,42,43,44\\
  &&&&&\quad ,45,46,47,48,4,5,6,7,8,9,10,11,12,13,14\\
  &&&&&\quad ,15,16\\
  5&2&new&0&cache&32,35,36,40,45\\
  5&2&gate&36&2Q&36,32\\
  5&2&gate&37&2Q&40,35\\
  5&2&gate&40&2Q&45,40\\
  5&2&new&0&cache&31,36,41,46\\
  5&2&gate&38&2Q&41,36\\
  5&2&gate&39&2Q&36,31\\
  5&2&gate&40&2Q&46,41\\
  5&2&new&0&cache&27,32,33,37,42\\
  5&2&gate&35&2Q&37,33\\
  5&2&gate&37&2Q&32,27\\
  5&2&gate&38&2Q&42,37\\
  5&2&gate&40&2Q&37,32\\
  5&2&new&0&cache&28,33,34,38,43\\
  5&2&gate&35&2Q&38,34\\
  5&2&gate&37&2Q&43,38\\
  5&2&gate&38&2Q&33,28\\
  5&2&gate&40&2Q&38,33\\
  5&2&new&0&cache&25,29,34,42,47\\
  5&2&gate&38&2Q&34,29\\
  5&2&gate&39&2Q&47,42\\
  5&2&gate&40&2Q&29,25\\
  5&2&new&0&cache&23,27,43,48\\
  5&2&gate&39&2Q&48,43\\
  5&2&gate&39&2Q&27,23\\
  5&2&new&0&cache&24,28,34,39\\
  5&2&gate&39&2Q&39,34\\
  5&2&gate&39&2Q&28,24\\
  5&2&write&0&disk&0,1,2,3,49,50,51,52\\
  \hline
\end{longtable}
}

\vskip2ex

The following listing is for the 54-qubit circuit.
{\scriptsize
  \begin{longtable}{|c|c|c|c|c|l|}
  \hline
  \hline
  Tensor&Phase&Mode&Depth&Gate&Arguments\\
  \hline
  \endfirsthead
  \hline
  Tensor&Phase&Mode&Depth&Gate&Arguments\\
  \hline
  \endhead
  \hline
  \endfoot
  \hline
  \hline
  \endlastfoot

  0&0&define&0&qubits&54\\
  0&0&define&0&disk&0,1,2,3,4,5,6,7,8,23,24,25,26,27,28\\
  &&&&&\quad ,29,30,31,50,51,52,53\\
  1&1&new&0&tensor&27,0,0,1,2,3,4,5,6,7,8,9,10,11,12\\
  &&&&&\quad ,13,14,15,16,17,18,19,20,21,22,23,24,25,26\\
  1&1&new&0&cache&10,14,18,19,23\\
  1&1&gate&1&2Q&14,10\\
  1&1&gate&1&2Q&23,19\\
  1&1&gate&3&2Q&18,14\\
  1&1&new&0&cache&2,6,10,18,23\\
  1&1&gate&2&2Q&6,2\\
  1&1&gate&4&2Q&10,6\\
  1&1&new&0&cache&1,5,6,18,23\\
  1&1&gate&2&2Q&5,1\\
  1&1&gate&6&2Q&6,1\\
  1&1&new&0&cache&5,9,14,18,23\\
  1&1&gate&3&2Q&9,5\\
  1&1&gate&6&2Q&14,9\\
  1&1&new&0&cache&0,5,18,23\\
  1&1&gate&5&2Q&5,0\\
  1&1&new&0&cache&18,22,23,26\\
  1&1&gate&2&2Q&26,22\\
  1&1&new&0&cache&18,20,23,24,26\\
  1&1&gate&1&2Q&24,20\\
  1&1&new&0&cache&12,16,18,20,23\\
  1&1&gate&2&2Q&16,12\\
  1&1&gate&4&2Q&20,16\\
  1&1&new&0&cache&18,20,21,23,25\\
  1&1&gate&2&2Q&25,21\\
  1&2&entgl&0&tensor&27,8,0,1,2,3,4,5,6,7,8,9,10,11,12\\
  &&&&&\quad ,13,14,15,16,17,18,19,20,21,22,23,24,25,26,-1\\
  &&&&&\quad ,-2,-3,-4,-5,-6,-7,-8,-9,-10,-11,-12,-13,-14,-15,-16\\
  1&2&entgl&3&EI&25,-1,-2\\
  1&2&entgl&3&EI&24,-3,-4\\
  1&2&entgl&4&EI&26,-5,-6\\
  1&2&entgl&4&EI&23,-7,-8\\
  1&2&new&0&cache&18,20,23,25\\
  1&2&gate&5&2Q&25,20\\
  1&2&gate&6&2Q&23,18\\
  1&2&new&0&cache&11,15,18,23,25\\
  1&2&gate&2&2Q&15,11\\
  1&2&new&0&cache&9,14,15,19,24\\
  1&2&gate&3&2Q&19,15\\
  1&2&gate&5&2Q&24,19\\
  1&2&gate&8&2Q&19,14\\
  1&2&gate&10&2Q&14,9\\
  1&3&entgl&0&tensor&28,13,0,1,2,3,4,5,6,7,8,9,10,11,12\\
  &&&&&\quad ,13,14,15,16,17,18,19,20,21,22,23,24,25,26,-1\\
  &&&&&\quad ,-2,-3,-4,-5,-6,-7,-8,-9,-10,-11,-12,-13,-14,-15,-16\\
  1&3&entgl&7&EI&24,-9,-10\\
  1&3&entgl&7&EI&23,-11,-12\\
  1&3&entgl&8&EI&25,-15,-16\\
  1&3&new&0&cache&14,18,19,23,24\\
  1&3&gate&9&2Q&24,19\\
  1&3&gate&10&2Q&23,18\\
  1&3&gate&12&2Q&19,14\\
  1&3&new&0&cache&0,5,10,15,20\\
  1&3&gate&6&2Q&15,10\\
  1&3&gate&7&2Q&20,15\\
  1&3&gate&8&2Q&10,5\\
  1&3&gate&9&2Q&5,0\\
  1&3&gate&10&2Q&15,10\\
  1&3&gate&12&2Q&10,5\\
  1&3&new&0&cache&10,14,18,20,25\\
  1&3&gate&9&2Q&25,20\\
  1&3&gate&13&2Q&14,10\\
  1&3&gate&15&2Q&18,14\\
  1&3&new&0&cache&13,15,17,20,21\\
  1&3&gate&1&2Q&17,13\\
  1&3&gate&4&2Q&21,17\\
  1&3&gate&11&2Q&20,15\\
  1&3&new&0&cache&3,7,11,21,26\\
  1&3&gate&1&2Q&7,3\\
  1&3&gate&4&2Q&11,7\\
  1&3&gate&6&2Q&26,21\\
  1&3&new&0&cache&1,5,6,11,16\\
  1&3&gate&5&2Q&16,11\\
  1&3&gate&8&2Q&11,6\\
  1&3&gate&10&2Q&6,1\\
  1&3&gate&14&2Q&5,1\\
  1&3&new&0&cache&0,1,5,9,26\\
  1&3&gate&15&2Q&9,5\\
  1&3&gate&18&2Q&5,1\\
  1&3&gate&19&2Q&9,5\\
  1&3&gate&21&2Q&5,0\\
  1&3&new&0&cache&6,11,15,16,21\\
  1&3&gate&7&2Q&21,16\\
  1&3&gate&9&2Q&16,11\\
  1&3&gate&12&2Q&11,6\\
  1&3&gate&14&2Q&15,11\\
  1&3&new&0&cache&3,4,8,12,13\\
  1&3&gate&1&2Q&8,4\\
  1&3&gate&3&2Q&12,8\\
  1&3&gate&5&2Q&8,3\\
  1&3&gate&7&2Q&13,8\\
  1&3&gate&9&2Q&8,3\\
  1&3&gate&11&2Q&13,8\\
  1&3&gate&13&2Q&8,4\\
  1&3&new&0&cache&2,6,7,12,17\\
  1&3&gate&5&2Q&17,12\\
  1&3&gate&6&2Q&7,2\\
  1&3&gate&7&2Q&12,7\\
  1&3&gate&10&2Q&7,2\\
  1&3&gate&14&2Q&6,2\\
  1&3&new&0&cache&6,9,10,14,18\\
  1&3&gate&16&2Q&10,6\\
  1&3&gate&17&2Q&14,10\\
  1&3&gate&19&2Q&18,14\\
  1&3&gate&22&2Q&14,9\\
  1&3&new&0&cache&1,2,6,10,26\\
  1&3&gate&18&2Q&6,2\\
  1&3&gate&20&2Q&10,6\\
  1&3&gate&22&2Q&6,1\\
  1&3&new&0&cache&7,12,13,17,22\\
  1&3&gate&8&2Q&22,17\\
  1&3&gate&9&2Q&17,12\\
  1&3&gate&11&2Q&12,7\\
  1&3&gate&12&2Q&22,17\\
  1&3&gate&13&2Q&17,13\\
  1&3&new&0&cache&3,7,11,21,26\\
  1&3&gate&13&2Q&7,3\\
  1&3&gate&16&2Q&11,7\\
  1&3&gate&17&2Q&7,3\\
  1&4&entgl&0&tensor&30,13,0,1,2,3,4,5,6,7,8,9,10,11,12\\
  &&&&&\quad ,13,14,15,16,17,18,19,20,21,22,23,24,25,26,-1\\
  &&&&&\quad ,-2,-3,-4,-5,-6,-7,-8,-9,-10,-11,-12,-13,-14,-15,-16\\
  1&4&entgl&8&EI&26,-13,-14\\
  1&4&new&0&cache&8,12,16,21,26\\
  1&4&gate&10&2Q&26,21\\
  1&4&gate&11&2Q&21,16\\
  1&4&gate&14&2Q&16,12\\
  1&4&gate&15&2Q&12,8\\
  1&4&new&0&cache&4,8\\
  1&4&gate&17&2Q&8,4\\
  2&5&new&0&tensor&27,2,27,28,29,30,31,32,33,34,35,36,37,38,39\\
  &&&&&\quad ,40,41,42,43,44,45,46,47,48,49,50,51,52,53,-1\\
  &&&&&\quad ,-2,-3,-4,-5,-6,-7,-8,-9,-10,-11,-12,-13,-14,-15,-16\\
  2&5&new&0&cache&37,41,45,46,50\\
  2&5&gate&1&2Q&50,46\\
  2&5&gate&2&2Q&41,37\\
  2&5&gate&3&2Q&45,41\\
  2&5&gate&6&2Q&50,45\\
  2&5&gate&10&2Q&50,45\\
  2&5&new&0&cache&31,35,38,42,46\\
  2&5&gate&2&2Q&35,31\\
  2&5&gate&2&2Q&42,38\\
  2&5&gate&4&2Q&46,42\\
  2&5&new&0&cache&30,34,35,39,43\\
  2&5&gate&1&2Q&34,30\\
  2&5&gate&1&2Q&43,39\\
  2&5&gate&3&2Q&39,35\\
  2&5&new&0&cache&30,31,35,40,44\\
  2&5&gate&1&2Q&44,40\\
  2&6&entgl&0&tensor&27,6,27,28,29,30,31,32,33,34,35,36,37,38,39\\
  &&&&&\quad ,40,41,42,43,44,45,46,47,48,49,50,51,52,53,-1\\
  &&&&&\quad ,-2,-3,-4,-5,-6,-7,-8,-9,-10,-11,-12,-13,-14,-15,-16\\
  2&6&entgl&4&E2Q&30,-5,-6\\
  2&6&entgl&4&E2Q&27,-7,-8\\
  2&6&entgl&8&E2Q&31,-13,-14\\
  2&6&new&0&cache&30,34,35,38,40\\
  2&6&gate&3&2Q&38,34\\
  2&6&gate&6&2Q&35,30\\
  2&6&gate&7&2Q&40,35\\
  2&6&new&0&cache&28,30,32,35,40\\
  2&6&gate&2&2Q&32,28\\
  2&6&new&0&cache&30,32,35,36,40\\
  2&6&gate&4&2Q&36,32\\
  2&6&new&0&cache&36,41,46,47,51\\
  2&6&gate&2&2Q&51,47\\
  2&6&gate&5&2Q&41,36\\
  2&6&gate&6&2Q&51,46\\
  2&6&gate&8&2Q&46,41\\
  2&6&gate&9&2Q&41,36\\
  2&6&gate&10&2Q&51,46\\
  2&6&gate&12&2Q&46,41\\
  2&6&new&0&cache&30,35,40,46,50\\
  2&6&gate&13&2Q&50,46\\
  2&6&new&0&cache&27,30,32,35,40\\
  2&6&gate&5&2Q&32,27\\
  2&6&new&0&cache&30,35,40,43,47\\
  2&6&gate&4&2Q&47,43\\
  2&6&new&0&cache&30,35,38,40,43\\
  2&6&gate&6&2Q&43,38\\
  2&7&entgl&0&tensor&27,10,27,28,29,30,31,32,33,34,35,36,37,38,39\\
  &&&&&\quad ,40,41,42,43,44,45,46,47,48,49,50,51,52,53,-1\\
  &&&&&\quad ,-2,-3,-4,-5,-6,-7,-8,-9,-10,-11,-12,-13,-14,-15,-16\\
  2&7&entgl&3&E2Q&28,-3,-4\\
  2&7&entgl&8&E2Q&30,-15,-16\\
  2&7&new&0&cache&29,30,33,35,40\\
  2&7&gate&1&2Q&33,29\\
  2&7&gate&10&2Q&35,30\\
  2&7&gate&11&2Q&40,35\\
  2&7&new&0&cache&27,32,33,37,42\\
  2&7&gate&4&2Q&37,33\\
  2&7&gate&6&2Q&42,37\\
  2&7&gate&7&2Q&37,32\\
  2&7&gate&9&2Q&32,27\\
  2&7&new&0&cache&37,42,47,48,52\\
  2&7&gate&2&2Q&52,48\\
  2&7&gate&5&2Q&52,47\\
  2&7&gate&8&2Q&47,42\\
  2&7&gate&9&2Q&52,47\\
  2&7&gate&10&2Q&42,37\\
  2&7&gate&12&2Q&47,42\\
  2&7&new&0&cache&29,32,37,41,45\\
  2&7&gate&11&2Q&37,32\\
  2&7&gate&14&2Q&41,37\\
  2&7&gate&15&2Q&45,41\\
  2&7&new&0&cache&28,33,38,47,51\\
  2&7&gate&5&2Q&33,28\\
  2&7&gate&8&2Q&38,33\\
  2&7&gate&14&2Q&51,47\\
  2&7&new&0&cache&29,34,39,44,48\\
  2&7&gate&3&2Q&48,44\\
  2&7&gate&5&2Q&44,39\\
  2&7&new&0&cache&38,43,48,49,53\\
  2&7&gate&1&2Q&53,49\\
  2&7&gate&5&2Q&53,48\\
  2&7&gate&7&2Q&48,43\\
  2&7&gate&9&2Q&53,48\\
  2&7&gate&10&2Q&43,38\\
  2&7&gate&11&2Q&48,43\\
  2&8&entgl&0&tensor&28,13,27,28,29,30,31,32,33,34,35,36,37,38,39\\
  &&&&&\quad ,40,41,42,43,44,45,46,47,48,49,50,51,52,53,-1\\
  &&&&&\quad ,-2,-3,-4,-5,-6,-7,-8,-9,-10,-11,-12,-13,-14,-15,-16\\
  2&8&entgl&3&E2Q&29,-1,-2\\
  2&8&entgl&7&E2Q&28,-11,-12\\
  2&8&new&0&cache&28,33,38,42,46\\
  2&8&gate&9&2Q&33,28\\
  2&8&gate&12&2Q&38,33\\
  2&8&gate&14&2Q&42,38\\
  2&8&gate&16&2Q&46,42\\
  2&8&new&0&cache&46,48,50,52\\
  2&8&gate&14&2Q&52,48\\
  2&8&gate&17&2Q&50,46\\
  2&8&new&0&cache&29,34,39,44,49\\
  2&8&gate&6&2Q&34,29\\
  2&8&gate&7&2Q&49,44\\
  2&8&gate&8&2Q&39,34\\
  2&8&gate&9&2Q&44,39\\
  2&8&gate&11&2Q&49,44\\
  2&8&new&0&cache&29,40,44,48,52\\
  2&8&gate&13&2Q&44,40\\
  2&8&gate&15&2Q&48,44\\
  2&8&gate&17&2Q&44,40\\
  2&8&gate&18&2Q&52,48\\
  2&8&gate&19&2Q&48,44\\
  2&8&new&0&cache&29,34,48,49,53\\
  2&8&gate&13&2Q&53,49\\
  2&8&gate&17&2Q&53,49\\
  2&8&gate&21&2Q&53,48\\
  2&9&entgl&0&tensor&30,13,27,28,29,30,31,32,33,34,35,36,37,38,39\\
  &&&&&\quad ,40,41,42,43,44,45,46,47,48,49,50,51,52,53,-1\\
  &&&&&\quad ,-2,-3,-4,-5,-6,-7,-8,-9,-10,-11,-12,-13,-14,-15,-16\\
  2&9&entgl&7&E2Q&29,-9,-10\\
  2&9&new&0&cache&29,34,39,43,47\\
  2&9&gate&10&2Q&34,29\\
  2&9&gate&12&2Q&39,34\\
  2&9&gate&13&2Q&43,39\\
  2&9&gate&16&2Q&47,43\\
  2&9&new&0&cache&47,51\\
  2&9&gate&18&2Q&51,47\\
  0&10&slice&0&disk&0,1,2,3,4,50,51,52,53\\
  1&10&all2all&0&tensor&32,6,-1,-2,-3,-4,-5,-6,-7,-8,-9,-10,-11,-12,-13\\
  &&&&&\quad ,-14,-15,-16,11,12,13,14,15,16,17,18,19,20,21,22\\
  &&&&&\quad ,23,24,25,26,5,6,7,8,9,10\\
  2&10&all2all&0&tensor&32,7,-1,-2,-3,-4,-5,-6,-7,-8,-9,-10,-11,-12,-13\\
  &&&&&\quad ,-14,-15,-16,27,28,29,30,31,32,33,34,35,36,37,38\\
  &&&&&\quad ,39,40,41,42,43,44,45,46,47,48,49\\
  3&1&new&0&tensor&32,13,11,12,13,14,15,16,17,18,19,20,21,22,23\\
  &&&&&\quad ,24,25,26,27,28,29,30,31,32,33,34,35,36,37,38\\
  &&&&&\quad ,39,40,41,42,5,6,7,8,9,10,43,44,45,46,47\\
  &&&&&\quad ,48,49\\
  3&1&new&0&cache&15,19,23,27,28\\
  3&1&gate&11&2Q&28,23\\
  3&1&gate&13&2Q&23,19\\
  3&1&gate&15&2Q&19,15\\
  3&1&gate&16&2Q&27,23\\
  3&1&gate&17&2Q&23,19\\
  3&1&gate&20&2Q&27,23\\
  3&1&new&0&cache&11,15,18,19,23\\
  3&1&gate&18&2Q&15,11\\
  3&1&gate&19&2Q&19,15\\
  3&1&gate&22&2Q&23,18\\
  3&1&new&0&cache&24,29,33,37,41\\
  3&1&gate&11&2Q&29,24\\
  3&1&gate&13&2Q&33,29\\
  3&1&gate&16&2Q&37,33\\
  3&1&gate&18&2Q&41,37\\
  3&1&new&0&cache&20,24,28,32,36\\
  3&1&gate&13&2Q&24,20\\
  3&1&gate&14&2Q&32,28\\
  3&1&gate&15&2Q&28,24\\
  3&1&gate&16&2Q&36,32\\
  3&1&gate&18&2Q&32,28\\
  3&1&gate&20&2Q&36,32\\
  3&1&new&0&cache&12,16,20,27,32\\
  3&1&gate&16&2Q&20,16\\
  3&1&gate&18&2Q&16,12\\
  3&1&gate&21&2Q&32,27\\
  3&1&new&0&cache&14,19,20,24,28\\
  3&1&gate&17&2Q&24,20\\
  3&1&gate&19&2Q&28,24\\
  3&1&gate&21&2Q&24,19\\
  3&1&gate&24&2Q&19,14\\
  3&1&new&0&cache&16,20,26,31,35\\
  3&1&gate&12&2Q&31,26\\
  3&1&gate&14&2Q&35,31\\
  3&1&gate&20&2Q&20,16\\
  3&1&new&0&cache&25,30,34,38,42\\
  3&1&gate&12&2Q&30,25\\
  3&1&gate&13&2Q&34,30\\
  3&1&gate&15&2Q&38,34\\
  3&1&gate&18&2Q&42,38\\
  3&1&new&0&cache&21,25,29,33,37\\
  3&1&gate&14&2Q&25,21\\
  3&1&gate&15&2Q&29,25\\
  3&1&gate&17&2Q&33,29\\
  3&1&gate&20&2Q&37,33\\
  3&1&new&0&cache&18,23,28,33\\
  3&1&gate&21&2Q&33,28\\
  3&1&gate&23&2Q&28,23\\
  3&1&gate&26&2Q&23,18\\
  3&1&new&0&cache&17,20,21,25,29\\
  3&1&gate&16&2Q&21,17\\
  3&1&gate&18&2Q&25,21\\
  3&1&gate&19&2Q&29,25\\
  3&1&gate&21&2Q&25,20\\
  3&1&new&0&cache&22,26,30,34,38\\
  3&1&gate&14&2Q&26,22\\
  3&1&gate&16&2Q&30,26\\
  3&1&gate&17&2Q&34,30\\
  3&1&gate&18&2Q&26,22\\
  3&1&gate&19&2Q&38,34\\
  3&1&gate&20&2Q&30,26\\
  3&1&new&0&cache&19,24,29,34\\
  3&1&gate&22&2Q&34,29\\
  3&1&gate&23&2Q&29,24\\
  3&1&gate&25&2Q&24,19\\
  3&1&new&0&cache&13,17,21,26\\
  3&1&gate&17&2Q&17,13\\
  3&1&gate&20&2Q&21,17\\
  3&1&gate&22&2Q&26,21\\
  3&1&new&0&cache&26,31,35,39\\
  3&1&gate&15&2Q&39,35\\
  3&1&gate&18&2Q&35,31\\
  3&1&gate&24&2Q&31,26\\
  3&2&all2all&0&tensor&32,13,18,19,20,21,22,23,24,25,26,27,28,29,30\\
  &&&&&\quad ,31,32,33,34,35,36,37,38,39,40,41,42,43,44,45\\
  &&&&&\quad ,46,47,48,49,5,6,7,8,9,10,11,12,13,14,15\\
  &&&&&\quad ,16,17\\
  3&2&new&0&cache&27,32,37,42,46\\
  3&2&gate&20&2Q&46,42\\
  3&2&gate&22&2Q&42,37\\
  3&2&gate&23&2Q&37,32\\
  3&2&gate&25&2Q&32,27\\
  3&2&new&0&cache&33,38,39,43,47\\
  3&2&gate&17&2Q&43,39\\
  3&2&gate&20&2Q&47,43\\
  3&2&gate&22&2Q&43,38\\
  3&2&gate&24&2Q&38,33\\
  3&2&new&0&cache&28,33,38,43,48\\
  3&2&gate&23&2Q&48,43\\
  3&2&gate&25&2Q&33,28\\
  3&2&gate&26&2Q&43,38\\
  3&2&gate&28&2Q&38,33\\
  3&2&new&0&cache&23,28,30,35,39\\
  3&2&gate&19&2Q&39,35\\
  3&2&gate&22&2Q&35,30\\
  3&2&gate&27&2Q&28,23\\
  3&2&new&0&cache&29,34,39,44,49\\
  3&2&gate&21&2Q&44,39\\
  3&2&gate&23&2Q&49,44\\
  3&2&gate&24&2Q&39,34\\
  3&2&gate&25&2Q&44,39\\
  3&2&gate&26&2Q&34,29\\
  3&2&gate&27&2Q&49,44\\
  3&2&gate&28&2Q&39,34\\
  3&2&new&0&cache&24,25,29,30,33\\
  3&2&gate&24&2Q&30,25\\
  3&2&gate&27&2Q&29,24\\
  3&2&gate&29&2Q&33,29\\
  3&2&new&0&cache&30,35,40,44\\
  3&2&gate&23&2Q&40,35\\
  3&2&gate&26&2Q&35,30\\
  3&2&gate&27&2Q&40,35\\
  3&2&gate&29&2Q&44,40\\
  3&2&new&0&cache&36,41,45\\
  3&2&gate&19&2Q&45,41\\
  3&2&gate&21&2Q&41,36\\
  3&3&all2all&0&tensor&32,13,5,6,7,8,9,10,11,12,13,14,15,16,17\\
  &&&&&\quad ,18,19,20,21,22,23,24,25,26,27,28,29,30,31,32\\
  &&&&&\quad ,33,34,35,36,37,38,39,40,41,42,43,44,45,46,47\\
  &&&&&\quad ,48,49\\
  3&3&new&0&cache&9,14,19,23,27\\
  3&3&gate&26&2Q&14,9\\
  3&3&gate&28&2Q&19,14\\
  3&3&gate&29&2Q&23,19\\
  3&3&gate&32&2Q&27,23\\
  3&3&new&0&cache&10,15,20,25,30\\
  3&3&gate&22&2Q&15,10\\
  3&3&gate&23&2Q&20,15\\
  3&3&gate&25&2Q&25,20\\
  3&3&gate&28&2Q&30,25\\
  3&3&new&0&cache&5,10,15,30,34\\
  3&3&gate&24&2Q&10,5\\
  3&3&gate&26&2Q&15,10\\
  3&3&gate&29&2Q&34,30\\
  3&3&new&0&cache&7,11,15,20,24\\
  3&3&gate&20&2Q&11,7\\
  3&3&gate&27&2Q&20,15\\
  3&3&gate&29&2Q&24,20\\
  3&3&new&0&cache&11,16,21,26,31\\
  3&3&gate&21&2Q&16,11\\
  3&3&gate&23&2Q&21,16\\
  3&3&gate&26&2Q&26,21\\
  3&3&gate&28&2Q&31,26\\
  3&3&new&0&cache&6,11,16,31,35\\
  3&3&gate&24&2Q&11,6\\
  3&3&gate&25&2Q&16,11\\
  3&3&gate&30&2Q&35,31\\
  3&3&new&0&cache&16,21,25,29\\
  3&3&gate&27&2Q&21,16\\
  3&3&gate&30&2Q&25,21\\
  3&3&gate&31&2Q&29,25\\
  3&3&new&0&cache&8,12,17,22\\
  3&3&gate&19&2Q&12,8\\
  3&3&gate&21&2Q&17,12\\
  3&3&gate&24&2Q&22,17\\
  3&3&write&0&disk&0,1,2,3,4,50,51,52,53\\
  4&1&read&0&disk&23,24,25,26,27,28,29,30,31\\
  4&1&new&0&tensor&32,13,0,1,2,3,4,5,6,7,8,9,10,11,12\\
  &&&&&\quad ,13,14,15,16,17,18,19,20,21,22,32,33,34,35,36\\
  &&&&&\quad ,37,38,39,40,41,42,43,44,45,46,47,48,49,50,51\\
  &&&&&\quad ,52,53\\
  4&1&new&0&cache&0,5,10,14,18\\
  4&1&gate&25&2Q&5,0\\
  4&1&gate&28&2Q&10,5\\
  4&1&gate&29&2Q&14,10\\
  4&1&gate&31&2Q&18,14\\
  4&1&new&0&cache&0,1,5,6,9\\
  4&1&gate&26&2Q&6,1\\
  4&1&gate&30&2Q&5,1\\
  4&1&gate&31&2Q&9,5\\
  4&1&gate&34&2Q&5,1\\
  4&1&gate&35&2Q&9,5\\
  4&1&gate&37&2Q&5,0\\
  4&1&new&0&cache&6,11,15,19\\
  4&1&gate&28&2Q&11,6\\
  4&1&gate&30&2Q&15,11\\
  4&1&gate&31&2Q&19,15\\
  4&1&new&0&cache&2,6,7,10,12\\
  4&1&gate&22&2Q&7,2\\
  4&1&gate&23&2Q&12,7\\
  4&1&gate&26&2Q&7,2\\
  4&1&gate&30&2Q&6,2\\
  4&1&gate&32&2Q&10,6\\
  4&1&gate&34&2Q&6,2\\
  4&1&new&0&cache&6,9,10,14,18\\
  4&1&gate&33&2Q&14,10\\
  4&1&gate&35&2Q&18,14\\
  4&1&gate&36&2Q&10,6\\
  4&1&gate&38&2Q&14,9\\
  4&1&new&0&cache&1,6,12,17,22\\
  4&1&gate&25&2Q&17,12\\
  4&1&gate&28&2Q&22,17\\
  4&1&gate&38&2Q&6,1\\
  4&1&new&0&cache&3,8,13,17,21\\
  4&1&gate&21&2Q&8,3\\
  4&1&gate&23&2Q&13,8\\
  4&1&gate&25&2Q&8,3\\
  4&1&gate&27&2Q&13,8\\
  4&1&gate&29&2Q&17,13\\
  4&1&gate&32&2Q&21,17\\
  4&1&gate&33&2Q&17,13\\
  4&1&new&0&cache&3,7,11,12,15\\
  4&1&gate&27&2Q&12,7\\
  4&1&gate&29&2Q&7,3\\
  4&1&gate&32&2Q&11,7\\
  4&1&gate&33&2Q&7,3\\
  4&1&gate&34&2Q&15,11\\
  4&1&gate&36&2Q&11,7\\
  4&1&new&0&cache&2,7,12,16,20\\
  4&1&gate&30&2Q&16,12\\
  4&1&gate&32&2Q&20,16\\
  4&1&gate&38&2Q&7,2\\
  4&1&new&0&cache&3,4,8,12,16\\
  4&1&gate&29&2Q&8,4\\
  4&1&gate&31&2Q&12,8\\
  4&1&gate&33&2Q&8,4\\
  4&1&gate&34&2Q&16,12\\
  4&1&gate&35&2Q&12,8\\
  4&1&gate&37&2Q&8,3\\
  4&1&new&0&cache&8,13\\
  4&1&gate&39&2Q&13,8\\
  4&2&all2all&0&tensor&32,13,13,14,15,16,17,18,19,20,21,22,32,33,34\\
  &&&&&\quad ,35,36,37,38,39,40,41,42,43,44,45,46,47,48,49\\
  &&&&&\quad ,50,51,52,53,0,1,2,3,4,5,6,7,8,9,10\\
  &&&&&\quad ,11,12\\
  4&2&new&0&cache&39,43,48,49,53\\
  4&2&gate&25&2Q&53,48\\
  4&2&gate&27&2Q&48,43\\
  4&2&gate&29&2Q&43,39\\
  4&2&gate&29&2Q&53,49\\
  4&2&gate&33&2Q&53,49\\
  4&2&new&0&cache&42,44,47,48,52\\
  4&2&gate&21&2Q&52,47\\
  4&2&gate&24&2Q&47,42\\
  4&2&gate&25&2Q&52,47\\
  4&2&gate&30&2Q&52,48\\
  4&2&gate&31&2Q&48,44\\
  4&2&gate&34&2Q&52,48\\
  4&2&new&0&cache&40,44,48,53\\
  4&2&gate&33&2Q&44,40\\
  4&2&gate&35&2Q&48,44\\
  4&2&gate&37&2Q&53,48\\
  4&2&new&0&cache&34,37,38,42,47\\
  4&2&gate&26&2Q&42,37\\
  4&2&gate&28&2Q&47,42\\
  4&2&gate&30&2Q&42,38\\
  4&2&gate&31&2Q&38,34\\
  4&2&new&0&cache&41,43,46,47,51\\
  4&2&gate&22&2Q&51,46\\
  4&2&gate&24&2Q&46,41\\
  4&2&gate&26&2Q&51,46\\
  4&2&gate&30&2Q&51,47\\
  4&2&gate&32&2Q&47,43\\
  4&2&gate&34&2Q&51,47\\
  4&2&new&0&cache&35,39,43,47,52\\
  4&2&gate&31&2Q&39,35\\
  4&2&gate&33&2Q&43,39\\
  4&2&gate&36&2Q&47,43\\
  4&2&gate&37&2Q&52,47\\
  4&2&new&0&cache&32,36,37,41,46\\
  4&2&gate&25&2Q&41,36\\
  4&2&gate&27&2Q&37,32\\
  4&2&gate&28&2Q&46,41\\
  4&2&gate&30&2Q&41,37\\
  4&2&new&0&cache&38,42,45,46,50\\
  4&2&gate&22&2Q&50,45\\
  4&2&gate&26&2Q&50,45\\
  4&2&gate&29&2Q&50,46\\
  4&2&gate&32&2Q&46,42\\
  4&2&gate&33&2Q&50,46\\
  4&2&gate&34&2Q&42,38\\
  4&2&gate&36&2Q&46,42\\
  4&2&new&0&cache&33,37,41,45,50\\
  4&2&gate&31&2Q&45,41\\
  4&2&gate&32&2Q&37,33\\
  4&2&gate&34&2Q&41,37\\
  4&2&gate&35&2Q&45,41\\
  4&2&gate&38&2Q&50,45\\
  4&2&new&0&cache&46,51\\
  4&2&gate&38&2Q&51,46\\
  4&2&write&0&disk&23,24,25,26,27,28,29,30,31\\
  5&1&read&0&disk&0,1,2,3,4,50,51,52,53\\
  5&1&new&0&tensor&32,13,5,6,7,8,9,10,11,12,13,14,15,16,17\\
  &&&&&\quad ,18,19,20,21,22,23,24,25,26,27,28,29,30,31,32\\
  &&&&&\quad ,33,34,35,36,37,38,39,40,41,42,43,44,45,46,47\\
  &&&&&\quad ,48,49\\
  5&1&new&0&cache&5,10,15,19,23\\
  5&1&gate&33&2Q&23,19\\
  5&1&gate&35&2Q&19,15\\
  5&1&gate&38&2Q&15,10\\
  5&1&gate&40&2Q&10,5\\
  5&1&new&0&cache&7,12,17,21,25\\
  5&1&gate&34&2Q&25,21\\
  5&1&gate&36&2Q&21,17\\
  5&1&gate&37&2Q&17,12\\
  5&1&gate&39&2Q&12,7\\
  5&1&new&0&cache&17,22,26,30,34\\
  5&1&gate&30&2Q&26,22\\
  5&1&gate&32&2Q&30,26\\
  5&1&gate&33&2Q&34,30\\
  5&1&gate&34&2Q&26,22\\
  5&1&gate&36&2Q&30,26\\
  5&1&gate&40&2Q&22,17\\
  5&1&new&0&cache&21,26,31,35\\
  5&1&gate&34&2Q&35,31\\
  5&1&gate&38&2Q&26,21\\
  5&1&gate&40&2Q&31,26\\
  5&1&new&0&cache&18,23,27,29,33\\
  5&1&gate&33&2Q&33,29\\
  5&1&gate&36&2Q&27,23\\
  5&1&gate&38&2Q&23,18\\
  5&1&new&0&cache&24,27,28,32,36\\
  5&1&gate&30&2Q&32,28\\
  5&1&gate&31&2Q&28,24\\
  5&1&gate&32&2Q&36,32\\
  5&1&gate&34&2Q&32,28\\
  5&1&gate&36&2Q&36,32\\
  5&1&gate&37&2Q&32,27\\
  5&1&new&0&cache&14,19,20,24,28\\
  5&1&gate&33&2Q&24,20\\
  5&1&gate&35&2Q&28,24\\
  5&1&gate&37&2Q&24,19\\
  5&1&gate&40&2Q&19,14\\
  5&1&new&0&cache&6,11,16,20,21\\
  5&1&gate&36&2Q&20,16\\
  5&1&gate&37&2Q&16,11\\
  5&1&gate&39&2Q&21,16\\
  5&1&gate&40&2Q&11,6\\
  5&1&new&0&cache&15,20,25,29\\
  5&1&gate&35&2Q&29,25\\
  5&1&gate&37&2Q&25,20\\
  5&1&gate&39&2Q&20,15\\
  5&2&all2all&0&tensor&32,13,18,19,20,21,22,23,24,25,26,27,28,29,30\\
  &&&&&\quad ,31,32,33,34,35,36,37,38,39,40,41,42,43,44,45\\
  &&&&&\quad ,46,47,48,49,5,6,7,8,9,10,11,12,13,14,15\\
  &&&&&\quad ,16,17\\
  5&2&new&0&cache&33,36,37,41,46\\
  5&2&gate&36&2Q&37,33\\
  5&2&gate&37&2Q&41,36\\
  5&2&gate&40&2Q&46,41\\
  5&2&new&0&cache&32,37,42,47\\
  5&2&gate&38&2Q&42,37\\
  5&2&gate&39&2Q&37,32\\
  5&2&gate&40&2Q&47,42\\
  5&2&new&0&cache&28,33,34,38,43\\
  5&2&gate&35&2Q&38,34\\
  5&2&gate&37&2Q&33,28\\
  5&2&gate&38&2Q&43,38\\
  5&2&gate&40&2Q&38,33\\
  5&2&new&0&cache&29,34,35,39,44\\
  5&2&gate&35&2Q&39,35\\
  5&2&gate&37&2Q&44,39\\
  5&2&gate&38&2Q&34,29\\
  5&2&gate&40&2Q&39,34\\
  5&2&new&0&cache&25,30,35,43,48\\
  5&2&gate&38&2Q&35,30\\
  5&2&gate&39&2Q&48,43\\
  5&2&gate&40&2Q&30,25\\
  5&2&new&0&cache&23,28,44,49\\
  5&2&gate&39&2Q&49,44\\
  5&2&gate&39&2Q&28,23\\
  5&2&new&0&cache&24,29,35,40\\
  5&2&gate&39&2Q&40,35\\
  5&2&gate&39&2Q&29,24\\
  5&2&write&0&disk&0,1,2,3,4,50,51,52,53\\
  \hline
\end{longtable}
}

\end{document}